\documentclass[longauth]{aa}  
\usepackage{graphicx}
\usepackage{txfonts}
\usepackage{hyperref}
\usepackage{subcaption}
\usepackage{placeins}
\usepackage{enumerate}
\usepackage{placeins}
\usepackage{multirow}
\usepackage[switch]{lineno}

\hypersetup{
    colorlinks=true,
    linkcolor=blue,
    filecolor=blue,      
    urlcolor=blue,
    citecolor=blue
    }

\newcommand{\linkorcid}[1]{\href{https://orcid.org/#1}{\includegraphics[width=8pt]{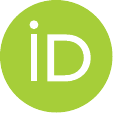}}}

\begin{document}

   \title{Photometric classification of supernovae detected by the Zwicky Transient Facility using noise augmentation}
   
   \titlerunning{Photometric classification of supernovae detected by ZTF}
   \authorrunning{A. Townsend et al.}

   \author{
   A.~Townsend\inst{\ref{hu}}\fnmsep\thanks{Corresponding author; alice.townsend@physik.hu-berlin.de}~\linkorcid{0000-0001-6343-3362} \and 
   J.~Nordin\inst{\ref{hu}} \linkorcid{0000-0001-8342-6274} \and
   M.~Kowalski\inst{\ref{hu},\ref{desy}}~\linkorcid{0000-0001-8594-8666} \and
   S.~Reusch\inst{\ref{aip}} \linkorcid{0000-0002-7788-628X} \and
   J.~P. Anderson\inst{\ref{eso}}~\linkorcid{0000-0003-0227-3451} \and
   E.~C.~Bellm\inst{\ref{dirac}}~\linkorcid{0000-0001-8018-5348} \and
   U.~Burgaz\inst{\ref{dublin}}~\linkorcid{0000-0003-0126-3999} \and
   T.~X.~Chen\inst{\ref{ipac}}~\linkorcid{0000-0001-9152-6224} \and
   T.-W.~Chen \inst{\ref{taiwan}}~\linkorcid{0000-0002-1066-6098} \and
   G.~Dimitriadis \inst{\ref{lancaster}}~\linkorcid{0000-0001-9494-179X} \and
   L.~Galbany \inst{\ref{ice},\ref{ieec}}~\linkorcid{0000-0002-1296-6887} \and
   A.~Goobar\inst{\ref{okc}}~\linkorcid{0000-0002-4163-4996} \and
   M.~J.~Graham\inst{\ref{caltech}}~\linkorcid{0000-0002-3168-0139} \and
   M.~Gromadzki\inst{\ref{poland}}~\linkorcid{0000-0002-1650-1518} \and
   C.~P.~Guti\'errez\inst{\ref{ice},\ref{ieec}}~\linkorcid{0000-0003-2375-2064} \and
   D.~Hale\inst{\ref{caltech_obs}} \and
   C.~Inserra\inst{\ref{cardiff}}~\linkorcid{0000-0002-3968-4409} \and
   M.~M.~Kasliwal\inst{\ref{caltech}}~\linkorcid{0000-0002-5619-4938} \and
   Y.-L.~Kim\inst{\ref{korea}}~\linkorcid{0000-0002-1031-0796} \and
   K.~Maguire\inst{\ref{dublin}}~\linkorcid{0000-0002-9770-3508} \and
   F.~J.~Masci\inst{\ref{ipac}}~\linkorcid{0000-0002-8532-9395} \and
   T.~E.~M\"uller-Bravo\inst{\ref{dublin},\ref{chile}}~\linkorcid{0000-0003-3939-7167} \and
   D.~A.~Perley\inst{\ref{liverpool}}~\linkorcid{0000-0001-8472-1996} \and
   R.~L.~Riddle\inst{\ref{ipac}}~\linkorcid{0000-0002-0387-370X} \and
   M.~Rigault\inst{\ref{lyon}}~\linkorcid{0000-0002-8121-2560} \and
   J.~van~Santen\inst{\ref{desy}}~\linkorcid{0000-0002-2412-9728} \and
   S.~Schulze\inst{\ref{northwestern}}~\linkorcid{0000-0001-6797-1889} \and
   M.~Smith\inst{\ref{lancaster}}~\linkorcid{0000-0002-3321-1432} \and
   J.~Sollerman\inst{\ref{stockholm}}~\linkorcid{0000-0003-1546-6615} \and
   S.~Yang\inst{\ref{china}}~\linkorcid{0000-0002-2898-6532}
          }

   \institute{Institut f\"ur Physik, Humboldt-Universit\"at zu Berlin, Newtonstr. 15, 12489 Berlin, Germany \label{hu}
     \and
     Deutsches Elektronen-Synchrotron, 15735 Zeuthen, Germany \label{desy}
     \and
     Leibniz-Institut für Astrophysik Potsdam (AIP), An der Sternwarte 16, 14482 Potsdam, Germany \label{aip}
     \and
     European Southern Observatory, Alonso de C\'ordova 3107, Casilla 19, Santiago, Chile \label{eso}
     \and
     DIRAC Institute, Department of Astronomy, University of Washington, 3910 15th Avenue NE, Seattle, WA 98195, USA \label{dirac}
     \and
     School of Physics, Trinity College Dublin, The University of Dublin, Dublin 2, Ireland \label{dublin}
     \and
     IPAC, California Institute of Technology, 1200 E. California Blvd, Pasadena, CA 91125, USA \label{ipac}
     \and
     Graduate Institute of Astronomy, National Central University, 300 Jhongda Road, 32001 Jhongli, Taiwan \label{taiwan}
     \and
     Department of Physics, Lancaster University, Lancs LA1 4YB, UK \label{lancaster}
     \and
     Institute of Space Sciences (ICE, CSIC), Campus UAB, Carrer de Can Magrans, s/n, E-08193 Barcelona, Spain \label{ice}
     \and
     Institut d'Estudis Espacials de Catalunya (IEEC), Edifici RDIT, Campus UPC, 08860 Castelldefels (Barcelona), Spain \label{ieec}
     \and
     Oskar Klein Centre, Department of Physics, Stockholm University, Albanova University Center, SE 106 91 Stockholm, Sweden \label{okc}
     \and
     Division of Physics, Mathematics and Astronomy, California Institute of Technology, Pasadena, CA 91125, USA \label{caltech}
     \and
     Astronomical Observatory, University of Warsaw, Al. Ujazdowskie 4, 00-478 Warszawa, Poland \label{poland}
     \and
     Caltech Optical Observatories, California Institute of Technology, Pasadena, CA  91125 \label{caltech_obs}
     \and
     Cardiff Hub for Astrophysics Research and Technology, School of Physics \& Astronomy, Cardiff University, Queens Buildings, The Parade, Cardiff, CF24 3AA, UK \label{cardiff}
     \and
     Department of Astronomy \& Center for Galaxy Evolution Research, Yonsei University, Seoul 03722, Republic of Korea \label{korea}
     \and
     Instituto de Ciencias Exactas y Naturales (ICEN), Universidad Arturo Prat, Chile \label{chile}
     \and
     Astrophysics Research Institute, Liverpool John Moores University, 146 Brownlow Hill, Liverpool L3 5RF, UK \label{liverpool}
     \and
     Univ Lyon, Univ Claude Bernard Lyon 1, CNRS, IP2I Lyon/IN2P3, UMR 5822, F-69622, Villeurbanne, France \label{lyon}
     \and
     Center for Interdisciplinary Exploration and Research in Astrophysics (CIERA), Northwestern University, 1800 Sherman Ave., Evanston, IL 60201, USA \label{northwestern}
     \and
     Department of Astronomy, Stockholm University, 10691 Stockholm, Sweden \label{stockholm}
     \and
     Henan Academy of Sciences, Zhengzhou 450046, Henan, China \label{china}
     }

   \date{Received; accepted}

 
  \abstract
   {Modern time-domain surveys, such as the Zwicky Transient Facility (ZTF), detect far more extragalactic transients than can be spectroscopically classified. Photometric classification offers a scalable alternative, enabling the identification of larger, fainter, and higher-redshift supernova samples suitable for applications such as Type Ia supernova (SN~Ia) cosmology. }
   {We present a feature-based photometric classifier for SNe detected by ZTF, with the primary goal of constructing a photometric SN~Ia sample for cosmological analyses. }
   {Our approach utilises the autoencoder architecture from the \texttt{ParSNIP} model \citep{Boone2021} to capture the intrinsic diversity of SN light curves. We trained the model on a spectroscopically classified ZTF SN sample, incorporating a realistic noise augmentation procedure that simulates the flux uncertainties of fainter sources. This enables the model to generalise to noisier, higher-redshift populations. Light curve features were used to train a gradient-boosted decision tree classifier, implemented in both binary (SN~Ia vs. non-Ia) and multi-class configurations. We validated our classifier on independent, fainter ZTF data with and without noise augmentation. To evaluate real-time performance, we also applied our classifier to live ZTF alerts and conducted a spectroscopic classification survey within the ePESSTO+ collaboration.}
   {We found that noise augmentation significantly improves classification performance, particularly for fainter sources. Our binary classifier achieves an SN~Ia recall of $(98.1 \pm 0.4)\%$, averaged across five train–test splits. SN~Ia recall exceeds 98\% for events with a peak apparent magnitude up to 20 and more than 10 detections, and remains above 96\% up to magnitude 20.5. Overall, 95\% of sources were correctly classified in both binary and multi-class modes. In our live classification survey, we correctly identified the class for 78\% of the targets, including rare events such as superluminous supernovae, despite a median of only nine detections per object.}
   {Our classifier performs efficiently on real ZTF data, including faint and noisy light curves. Applied to the full ZTF transient sample, it will enable construction of a large photometric SN~Ia dataset for cosmology. It also provides a practical tool for real-time spectroscopic target prioritisation, which will be essential for large-scale surveys such as the Vera C. Rubin Observatory’s Legacy Survey of Space and Time.}

   \keywords{supernovae: general --
                Methods: data analysis
               }

   \maketitle
%

\section{Introduction}
\label{section:intro}

In the era of wide-field, time-domain surveys such as the Zwicky Transient Facility \citep[ZTF;][]{Bellm2019a,Graham2019}, significantly more extragalactic transients are detected than can be spectroscopically classified with available resources. The volume of data is expected to increase dramatically with the arrival of deeper and more sensitive surveys, such as the Vera C. Rubin Observatory’s Legacy Survey of Space and Time \citep[LSST;][]{Ivezic2019}. This has driven the development of accurate, scalable, and robust machine learning classifiers capable of identifying and classifying transients using photometric data alone, with minimal reliance on spectroscopic follow-up.

A variety of classification approaches have emerged, ranging from feature-based methods \citep[e.g.,][]{Lochner2016,Villar2019,Boone2019,Miranda2022,deSoto2024} and template-fitting methods \citep[e.g.,][]{Sako2011} to deep learning architectures \citep[e.g.,][]{Muthukrishna2019a, Moller2020,Boone2021,Qu2021, Pimentel2023}, semi-supervised learning \citep[e.g.,][]{Richards2012,Villar2020}, transfer learning \citep[e.g.,][]{ZhangG2024}, and active learning \citep[e.g.,][]{Kennamer2020,Leoni2022}. Some classifiers are specialised to find rare transients such as superluminous supernovae (SLSNe) or tidal disruption events (TDEs) \citep[e.g.,][]{Gomez2020,Sheng2024,Stein2024}. While most classifiers rely exclusively on the photometric data of the transient itself, some have been developed that use only information from the host galaxy \citep[e.g.,][]{Gagliano2021,Kisley2023}.

Real-time brokers and pipelines such as the Automatic Learning for the Rapid Classification of Events Alert Broker \citep[ALeRCE;][]{Forster2021} and Fink \citep{Leoni2022} have demonstrated the practical utility of these models in live survey operations. Several additional classifiers have been developed to enable rapid detection and classification of transients in real-time surveys, using limited early-time photometric data. \citep[e.g.,][]{Muthukrishna2019a,Miranda2022,Gagliano2023,ZhangL2024a}. Furthermore, classification challenges that simulate live surveys continue to encourage innovation in the community, such as the Supernova Photometric Classification Challenge \citep[SPCC, ][]{Kessler2010} and the LSST data challenges with PLAsTiCC and ELAsTiCC simulations \citep{Hlozek2023,Narayan2023,Shah2025,Nordin2025}. 

One persistent issue in the field is the small and biased sample of spectroscopically confirmed supernovae available for model training. It is well established that machine learning models trained on samples that are not representative of the test sample tend to perform poorly. This limitation was highlighted in the SPCC \citep{Kessler2010,Richards2012,Lochner2016}, where the simulated training set was intentionally biased toward bright, low-redshift events typical of spectroscopic follow-up campaigns. Some classifiers addressed this challenge by augmenting the training set using Gaussian Process \citep[GP;][]{Rasmussen2006} models fitted to the original light curves. These GP-based methods generate additional synthetic light curves to better populate underrepresented regions of feature space, improving classification performance on the simulated test sets \citep{Revsbech2018,Pasquet2019,Boone2019}. \citet{Pasquet2019} and \citet{Boone2019} also introduced additional forms of noise augmentation, either by adding noise to the flux values or by resampling the redshift and scaling the flux accordingly. Both studies included further augmentation strategies such as randomly dropping observations or applying random time shifts. Further analyses using simulated LSST data have reinforced the effectiveness of this augmentation approach \citep{Carrick2021,Alves2022}. Although these results are promising, so far they are only validated on simulated datasets. Studies have shown that a model solely trained on simulated data performs poorly when classifying real data \citep[e.g.,][]{Revsbech2018}.

In a different approach, \citet{Boone2021} proposed to solve the issue of non-representative datasets with \texttt{ParSNIP}\footnote{\href{https://github.com/LSSTDESC/parsnip}{\texttt{github.com/LSSTDESC/parsnip}}}, a deep learning model designed to be invariant to observing conditions and to the transient distance. The model learns a latent representation of each light curve by explicitly disentangling the intrinsic features of the transient (such as its brightness evolution and colour) from observational factors such as redshift, host galaxy extinction, and survey cadence. This allows the model to generalise more effectively across the biased datasets typical of real surveys. \texttt{ParSNIP} has previously been utilised to classify a photometric SN sample for the Young Supernova Experiment Data Release 1 \citep[YSE DR1;][]{Aleo2023} and to make predictions for photometric classification using simulated Roman Space Telescope data \citep{Abdelhadi2024}.

Currently, the majority of photometric classifiers are trained and tested on simulated datasets to compensate for the lack of real, labelled data. However, these simulations, which are generated from transient template models, often fail to capture the full diversity of real transients. In this work, we extend the previous studies on noise augmentation by applying a new methodology to real data from ZTF. We have combined the \texttt{ParSNIP} model of \citet{Boone2021} with our noise augmentation algorithm to demonstrate that we can accurately classify the noisier and more sparsely sampled candidate supernovae found in the ZTF archive.

\subsection{The application of photometric classifiers in supernova cosmology}
\label{subsection:intro_snia}

Type Ia supernovae (SNe~Ia) are precise cosmological distance indicators due to their nearly uniform peak luminosities, enabling their use as standardisable candles \citep[e.g.,][]{Phillips1993,Riess1996,Tripp1998}. This approach led to the discovery of the accelerating expansion of the universe, attributed to dark energy \citep{Riess1998,Perlmutter1999}. Today, SNe~Ia remain essential for constraining key cosmological parameters such as the Hubble constant ($H_0$) and the dark energy equation-of-state parameter ($w$).

Since the discovery of cosmic acceleration, based on a sample of approximately 40 SNe~Ia, progressively larger datasets have been employed in cosmological analyses. One of the most comprehensive of these is the Pantheon+ compilation, which includes 1550 spectroscopically confirmed SNe~Ia \citep{Brout2022}. The ZTF SN~Ia data release 2 \citep[DR2;][]{Rigault2025} provides a publicly available sample of 3628 spectroscopically classified SNe~Ia, which is the largest release of SNe~Ia to date and increases the sample of low redshift SNe~Ia by an order of magnitude. In recent years, analyses with samples of photometrically classified SNe~Ia -- such as those from Pan-STARRS \citep[PS1;][]{Jones2018}, the Sloan Digital Sky Survey \citep[SDSS;][]{Sako2011,Campbell2013}  and combined datasets \citep{Popovic2024} -- have demonstrated that unbiased cosmological parameters can be recovered, in agreement with those derived from spectroscopically confirmed samples.

A notable example is the Dark Energy Survey \citep[DES;][]{Abbott2019,DES2024}, which recently released cosmological constraints based on a sample of 1635 photometrically identified SNe~Ia \citep{Moller2022,Vincenzi2024}. DES used \texttt{SuperNNova} \citep{Moller2020} as the baseline photometric classifier, supplemented with cross-checks using alternative methods such as \texttt{SCONE} \citep{Qu2021}. Simulations suggest a classification accuracy and purity exceeding 98\% \citep{Vincenzi2021,Vincenzi2023}. Crucially, the impact of photometric misclassification was found to be small, with uncertainties from photometric classification contributing less than 10\% of the total systematic error budget. 

To incorporate classification uncertainties into the cosmological analysis, both DES and PS1 adopted probabilistic frameworks. These analyses used the Bayesian Estimation Applied to Multiple Species \citep[BEAMS;][]{Kunz2013,Hlozek2012} approach or its extension, BEAMS with Bias Corrections \citep[BBC;][]{Kessler2017}. BEAMS marginalises over the probability that each event is an SN~Ia, thereby accounting for contamination from core-collapse SNe. The BBC extension applies corrections to the observed distance moduli, accounting for selection effects, contamination, and light curve fitting biases, all derived from realistic simulations. This enables the construction of a bias-corrected Hubble diagram from photometrically classified SNe~Ia.

The results of these studies indicate that contamination from other SN types in photometrically classified samples can be limited to approximately 5\% \citep[e.g.,][]{Jones2018,Vincenzi2021,Vincenzi2023,Moller2022}. At this level, contamination is a subdominant source of uncertainty, and cosmological constraints remain comparable to those from spectroscopic samples if classification probabilities and selection effects are correctly modelled. As a result, improving SN~Ia classification completeness (also known as recall) is key to increasing the statistical power of photometric samples. In the following analysis, we focus on SN~Ia recall as the primary classifier performance metric.

In this work, we present a feature-based photometric classifier for SNe detected by ZTF, with the primary goal of constructing a photometric SN~Ia sample for cosmological analyses. Our approach utilises the autoencoder architecture from the \texttt{ParSNIP} model \citep{Boone2021} to capture the intrinsic diversity of SN light curves. We trained the model on a spectroscopically classified ZTF SN sample, incorporating a realistic noise augmentation procedure that simulates the flux uncertainties of fainter sources. This enables the model to generalise to noisier, higher-redshift populations. Light curve features were used to train a gradient-boosted decision tree classifier, implemented in both binary (SN~Ia vs. non-Ia) and multi-class configurations. We validated our classifier on independent, fainter ZTF data with and without noise augmentation. To evaluate real-time performance, we also applied our classifier to live ZTF alerts and conducted a spectroscopic classification survey within the extended Public ESO Spectroscopic Survey for Transient Objects  \citep[ePESSTO+;][]{Smartt2015} at the 3.5-metre ESO New Technology Telescope (NTT).

The structure of this paper is as follows. Section~\ref{section:data} summarises the ZTF data used in this analysis, the transient classes considered, and how we divided the training and test samples. In Sect.~\ref{section:method}, we describe how we modelled the ZTF flux uncertainties to simulate higher-redshift sources and detail the noise augmentation procedure used to improve classification performance. Section~\ref{section:classification_method} outlines the generation of training and test sets, the training of the \texttt{ParSNIP} model, and the creation of the classifiers. In Sect.~\ref{section:results}, we present the classification results, discuss the optimisation of the augmentation parameters, and highlight the performance of one of our best models. We further evaluate the classifier on additional datasets, including the ZTF SN~Ia DR2 sample (Sect.~\ref{subsection:results_dr2}) and peculiar or uncommon subtypes initially excluded (Sect.~\ref{subsection:results_contam_pec}). We then discuss applications of our method to SN~Ia cosmology (Sect.~\ref{subsection:results_cosmo}) and real-time transient classification (Sect.~\ref{subsection:results_livetest}), which includes results from a live classification survey conducted with the ePESSTO+ collaboration. Finally, Sect.~\ref{section:conc} summarises our key findings and outlines the future steps toward constructing a photometrically classified ZTF SN sample.

\section{Data}
\label{section:data}

ZTF is an optical, wide-field, survey designed for time-domain astronomy \citep{Bellm2019a,Graham2019}. It operates using a 47-square-degree field-of-view camera mounted on the 48-inch telescope at Palomar Observatory (P48), which observes in three photometric bands: $g$, $r$ and $i$ \citep{Dekany2020}. The observing strategy includes a publicly available survey in $g$- and $r$-bands that surveys the northern sky every two days, alongside a higher-cadence partnership survey that includes $i$-band observations over a more limited area. The public survey initially operated on a three-day cadence during Phase I but transitioned to a two-day cadence thereafter \citep{Bellm2019b}. This analysis makes use of both the public and partnership datasets up to September 2024. On a typical full night, ZTF generates between 600,000 and 1.2 million alerts \citep{Masci2019,Patterson2019}. As a result, ZTF is a powerful tool for discovering and monitoring transients, making it particularly well-suited for building large, homogeneous samples of SNe essential for training machine learning models.

The Bright Transient Survey \citep[BTS;][]{Fremling2020,Perley2020} is a dedicated program within ZTF that aims to spectroscopically classify all extragalactic transients, excluding active galactic nuclei (AGN), with a peak brighter than 18.5 magnitude. This is achieved using low-resolution spectra from the Spectral Energy Distribution Machine (SEDM) that operates on the Palomar 60-inch telescope \citep{Blagorodnova2018,Rigault2019,Kim2022}. This constitutes the majority of the classified sources in the ZTF dataset, and is the largest brightness-limited SN survey to date. As well as relying on human scanners to request follow-up, BTS employs several machine learning classifiers to automate classification. SNIascore \citep{Fremling2021} specialises in real-time classification of SNe~Ia from SEDM spectra, achieving <0.6\% false-positive rates. CCSNscore \citep{Sharma2025} extends this framework to core-collapse SNe, whilst also incorporating ZTF light curves, to distinguish SNe~II from SNe~Ib/c with high accuracy. BTSbot \citep{Rehemtulla2024} is a discovery-to-classification pipeline that autonomously triggers follow-up observations and processes data through these classifiers, enabling automated reporting of ZTF SNe.

The majority of our dataset consisted of the BTS SN sample\footnote{This will be released in Qin et al. (in prep.)} collected up to September 2024, although additional ZTF objects were included to increase the diversity in our SLSN class (see Sect.~\ref{subsection:classes} for details). We obtained the list of classified transients and their corresponding classifications, redshifts, right ascensions, and declinations using the internal BTS Sample Explorer\footnote{A public version of the catalogue is available at:  \href{https://sites.astro.caltech.edu/ztf/bts/explorer.php}{\texttt{sites.astro.caltech.edu/ztf/bts/explorer.php}}}, selecting all classified supernovae regardless of peak magnitude. The forced photometry (FP) for our sample was acquired using the IPAC ZTF forced photometry service \citep{Masci2023}, including both public and partnership data. We then processed the data to remove the baseline (i.e. the background flux level at the transient’s location) using the alert processing software \texttt{AMPEL} \citep{Nordin2019}\footnote{\href{https://github.com/AmpelAstro/Ampel-ZTF/blob/master/ampel/ztf/alert/ZTFIPACForcedPhotometryAlertSupplier.py}{\texttt{ZTFIPACForcedPhotometryAlertSupplier.py}}}. Transient light that may be present in the reference image is largely removed during the baseline subtraction. We also removed unreliable data points from the FP light curves using warning flags from the IPAC service, which indicate issues with the science image, poor seeing, or baseline calculation.

\subsection{Class breakdown}
\label{subsection:classes}
We trained our model on the following classes of SNe: SN~Ia, SN~Ib/c, SN~II, SN~IIn, and SLSN. Table~\ref{table:train_classes} lists how individual subtypes were grouped into these classes, along with their sample sizes. Table~\ref{table:exlc_classes} shows the SN subtypes that were excluded from the model training. We included 91T-like SNe~Ia in our SN~Ia sample because, although they are slightly more luminous than normal SNe~Ia, they are generally well modelled by standard SN~Ia light curve models, and their peculiarity is primarily spectroscopic \citep{Dimitriadis2025}.

We note that some peculiar SNe~Ia are likely still present in our training sample due to incomplete or ambiguous subtyping, which may affect classification performance. The redshifts and classifications are preliminary, as they come from an internal database, and are being finalised within the BTS working group (Qin et al., in prep.). Additionally, there are updated SN~Ia subtypes presented in the ZTF SN~Ia DR2 publications \citep{Rigault2025,Dimitriadis2025,Burgaz2025}, which will be incorporated in future work.

We excluded both peculiar and uncommon transient subtypes from our model training due to insufficient sample sizes to reliably model their behaviour, which could otherwise bias the classifier. For example, SNe~IIb are known to be photometrically distinct from other SNe~II \citep[e.g.,][]{Pessi2019}. TDEs were also excluded because of their limited sample size. Additionally, SNe~Ia with the “Ia?” label in the BTS internal catalogue, indicating uncertain classification, were removed. We also expect that K-corrections (see Sect.~\ref{subsection:kcorr}) would be less accurate for the peculiar subtypes due to their different spectral properties. However, we discuss how such excluded events would be classified by our model in Sect.~\ref{subsection:results_contam_pec}.

Furthermore, we incorporated an additional sample of SLSNe to increase the diversity of the training sample, since the original BTS sample only contained 154 objects. We included classified events from \citet{Chen2023} for SLSNe-I and \citet{Kangas2022} and \citet{Pessi2025} for SLSNe-II. We also queried \texttt{Fritz}, a \texttt{SkyPortal} platform for members of the ZTF collaboration \citep{vanderWalt2019,Coughlin2023}, for additional SLSN candidates not included in the previous ZTF publications. This increased our SLSN sample by 84 objects.

\begin{table}[h]
\caption{\label{table:train_classes} Classes included from the BTS sample and additional SLSN samples, and their corresponding SN subtypes.}
\centering
\begin{tabular}{c | c | c | c}
\hline\hline
Class & Class total (\%) & Subtypes & Subtypes total \\ 
\hline
SN~Ia & 6614 (69.4\%) & Ia and Ia-norm & 6396 \\
 && Ia-91T & 218 \\
SN Ib/c & 551 (5.8\%) & Ib & 220 \\
 && Ib/c & 43 \\
 && Ic & 213 \\
 && Ic-BL & 75 \\
SN II & 1859 (19.5\%) & II and II-norm & 1684 \\
 && IIL & 2 \\
 && IIP & 173 \\
SLSN & 238 (2.5\%) & SLSN-I & 155 \\
 && SLSN-II & 79 \\
 && SLSN-I.5 & 4 \\
 SN IIn & 271 (2.8\%) & IIn & 271 \\
\hline
\end{tabular}
\end{table}

\begin{table}[h]
\caption{\label{table:exlc_classes} SN subtypes excluded in training.}
\centering
\begin{tabular}{c | c | c | c}
\hline\hline
Class & Class total & Subtypes & Subtypes total \\ 
\hline
SN~Ia & 254 & Ia? & 35 \\
 && Ia-pec & 41 \\
&& Ia-CSM & 28 \\
 && Ia-91bg & 86 \\
 && Ia-99aa & 1 \\
 && Ia-00cx & 3 \\
 && Iax (Ia-02xc) & 38 \\
 && Ia-03fg & 21 \\
 && Ia-18byg & 1 \\
SN Ib/c & 42 & Ib-pec & 4 \\
 && Ibn & 36 \\
 && Icn & 2 \\
SN II & 163 & IIb & 160 \\
 && IIb-pec & 1 \\
 && II-pec & 2 \\
TDE & 42 & TDE & 42 \\
\hline
\end{tabular}
\end{table}

\subsection{Training and test samples}
\label{subsection:traintest}
We trained our models using a 90\% training and 10\% test split. To assess model performance across different magnitude regimes, we divided the dataset into the following subsamples.
\subsubsection{\texttt{BrightZTF} and \texttt{FaintZTF}}
\label{subsubsection:brightfaintZTF}
\begin{figure}[t!]
\centering
\includegraphics[width=\hsize]{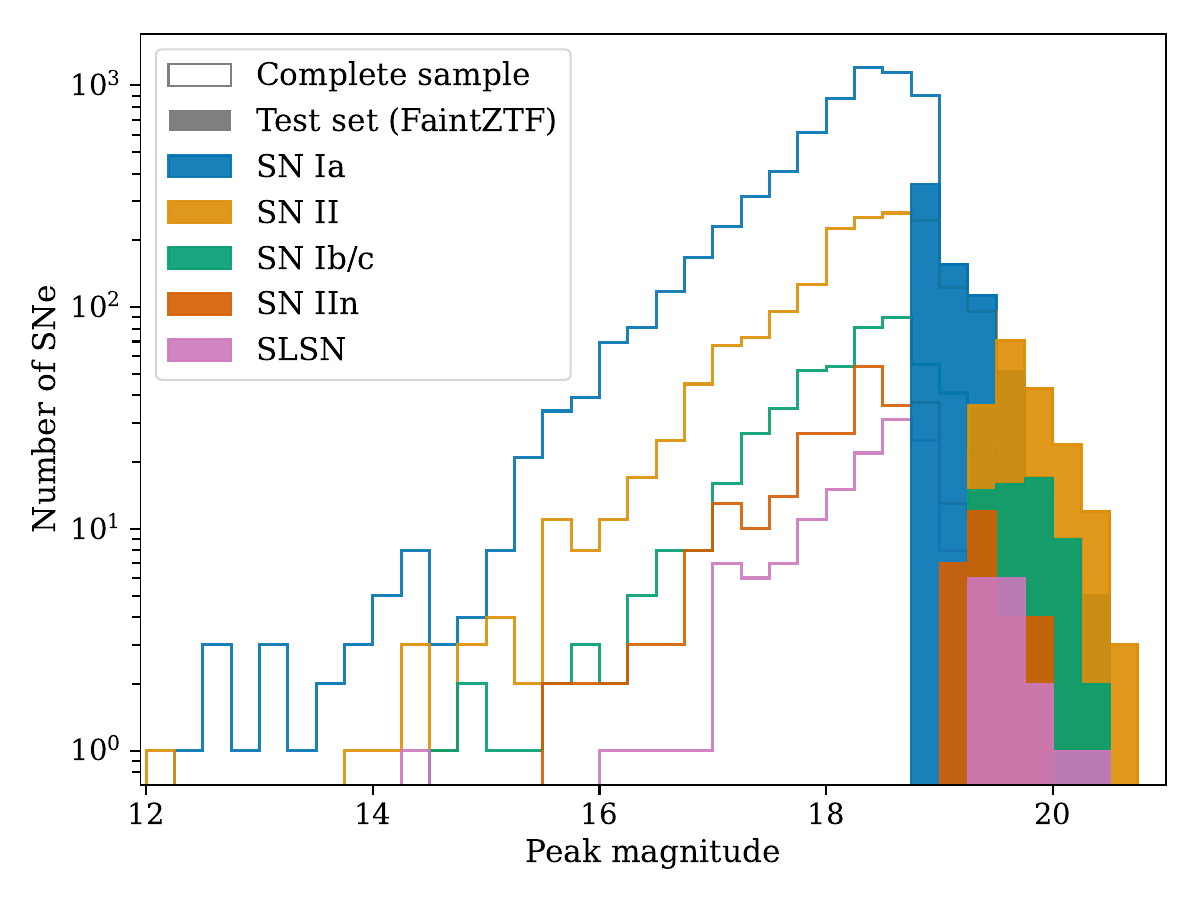}
    \caption{The peak apparent magnitude distribution for our sample, divided into classes of SN~Ia (blue), SN~II (orange), SN~Ib/c (green), SN~IIn (red), SLSN (pink). The filled histograms represent the test sample, \texttt{FaintZTF}, and the outline represents the full dataset. The peak apparent magnitude corresponds to the brightest detection in any ZTF band ($g$, $r$ or $i$).}
    \label{figure:faintZTF}
\end{figure}

We split the data by class (as defined in Table~\ref{table:train_classes}) into the brightest 90\% -- referred to as \texttt{BrightZTF} -- and the faintest 10\% -- referred to as \texttt{FaintZTF} -- based on peak apparent magnitude (using the brightest detection in any band). This is illustrated by the peak apparent magnitude distribution shown in Fig.~\ref{figure:faintZTF}, where the filled histograms represent \texttt{FaintZTF} and the outline shows the complete dataset. In practice, \texttt{FaintZTF} is utilised as the test set, and additional quality cuts (described in Sect.~\ref{subsection:create_train}) are applied before we perform the classifications. This setup provides a benchmark for evaluating the effectiveness of our data augmentation in improving classifier performance at fainter magnitudes. When augmentation is applied to these samples, we refer to them as \texttt{BrightZTF+Noise} and \texttt{FaintZTF+Noise}.

\subsubsection{\texttt{RandomZTF}}
\label{subsubsection:randomZTF}
To evaluate performance across a broader magnitude range, we also perform a random split of the data -- referred to as \texttt{RandomZTF}. In this case, we generate five different training and test sets using five random seeds, and report the average classification score. When augmentation is applied to these samples, we refer to them as \texttt{RandomZTF+Noise}.

\section{Augmentation method}
\label{section:method}
To create a training sample that is more representative of the test sample, we implemented an augmentation method that generates fainter, noisier synthetic light curves. This is similar to the methodology of \citet{Boone2019}, where redshifts are resampled, except we did not include Gaussian Process fitting to our light curves. Instead, we accurately modelled the behaviour of the ZTF flux errors in different magnitude regimes and scaled our flux and flux error values accordingly (Sect.~\ref{subsection:errmodel}). We also implemented K-corrections (Sect.~\ref{subsection:kcorr}) and additional augmentation methods such as removing data points (Sect.~\ref{subsection:noisify}).

\subsection{Modelling the ZTF flux errors at higher redshifts}
\label{subsection:errmodel}
To accurately generate the errors for the augmented light curves, we determined an error model for the ZTF data by attempting to simulate the flux errors from the original sample of ZTF light curves. We start by assuming that the dominant contributions to the true flux error, $\sigma_\mathrm{t}$, are due to the Poisson counting of the photons ($\sigma_\mathrm{t} \propto \sqrt{f_t}$, where $f_\mathrm{t}$ is the true flux) and a bias term that is dominant at lower flux values, $e_\mathrm{b}$ (where $\sigma_\mathrm{t} \propto e_\mathrm{b}$). Then, the variance of the true flux error would be
\begin{equation}
\label{equation:errmodel_basic}
    \sigma_\mathrm{t}^2 = \:f_\mathrm{t} + e_\mathrm{b}^2.
\end{equation}
When comparing this with the ZTF data, we found that this does not capture the true nature of the noise, particularly at lower flux values. In ZTF photometry, the actual noise affecting a measurement arises from a combination of statistical, systematic, and environmental factors, not just Poisson noise from the source photons. One important component is sky background noise, which results from Poisson fluctuations in the sky brightness, including contributions from moonlight. Because this background is integrated within the PSF aperture, it adds a flux-dependent noise term that varies with observing conditions (e.g., lunation and airmass). Calibration issues such as zeropoint drift, PSF mismatches, and astrometric misalignments also introduce uncertainties that are not purely random, but often correlated in space, time, or observing conditions. These effects are further compounded in difference imaging, particularly for transients in bright host galaxies, where host light may be incompletely subtracted, leaving behind spatially variable residuals that contaminate the transient signal. While the noise properties can vary somewhat between filters due to different throughput and sky brightness levels, we adopt a single model across all three ZTF bands ($g$, $r$, $i$) as a simplification, since the dominant sources of noise exhibit similar qualitative behaviour across filters. Overall, these contributions combine to produce measurement uncertainties that are neither strictly Poissonian nor entirely uncorrelated, motivating the use of an empirical noise model to approximate the observed errors in real ZTF light curves.

To account for these effects, we introduce two empirical fitting terms: a random exponential component, $\epsilon$, and a constant offset, $\delta$. This results in the following relationship between flux and flux error:
\begin{equation}
\label{equation:errmodel}
    \frac{\sigma_\mathrm{t}}{f_\mathrm{t}} = \sqrt{\frac{1}{f_\mathrm{t} }+ \frac{e_\mathrm{b}^2}{f_\mathrm{t}^2}} + \epsilon + \delta, \qquad \epsilon \sim \text{Exp}(\lambda_f),
\end{equation}
where the rate parameter of the exponential distribution is linearly dependent on the flux such that \mbox{$\lambda_f = m \cdot f_\mathrm{t} + c$}.

\begin{figure*}[htbp]
\centering
\includegraphics[width=0.8\textwidth]{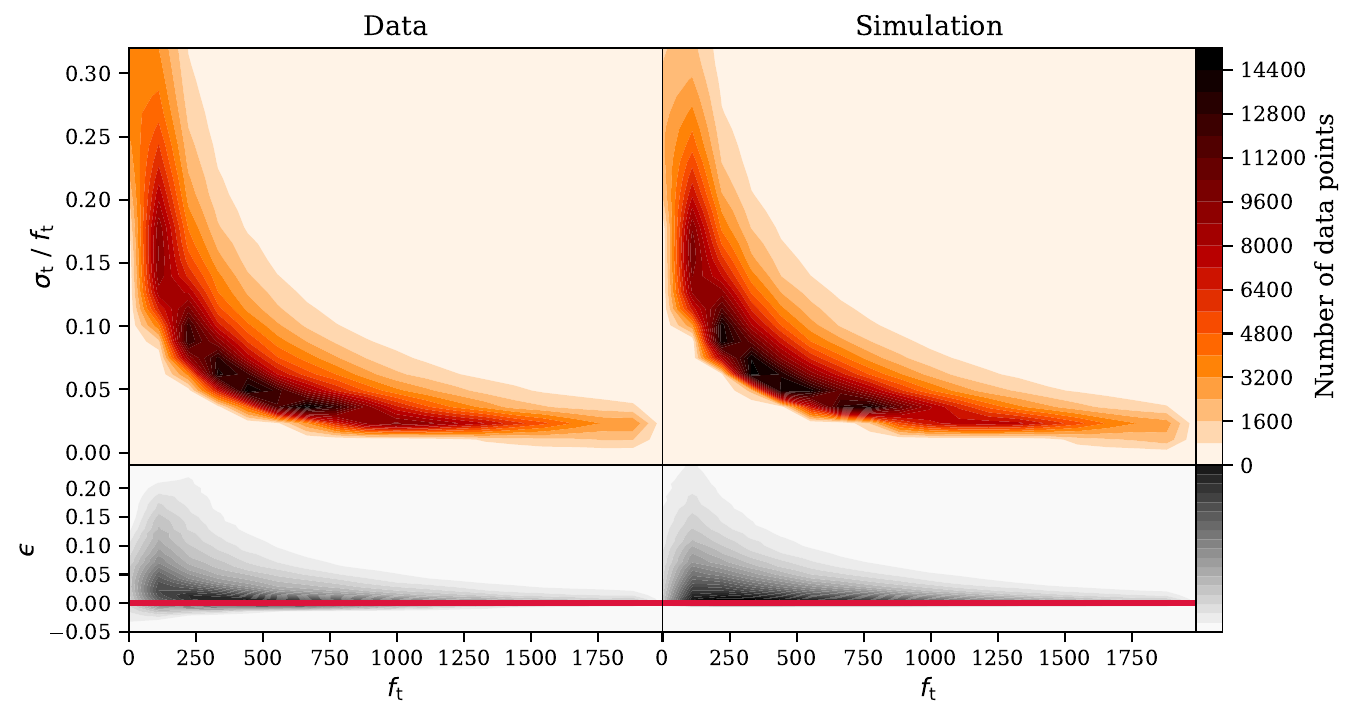}
    \caption{Comparison of the $\sigma_\mathrm{t}/f_\mathrm{t}$ vs. $f_\mathrm{t}$ distributions of the original ZTF light curves (upper left) and the ZTF light curves with flux errors simulated according to Equation~\ref{equation:errmodel} (upper right). The corresponding residual distributions for $\epsilon$ are shown in the lower panels.}
    \label{figure:error_model}
\end{figure*}

\begin{figure*}[htbp]
\centering
\includegraphics[width=0.85\textwidth]{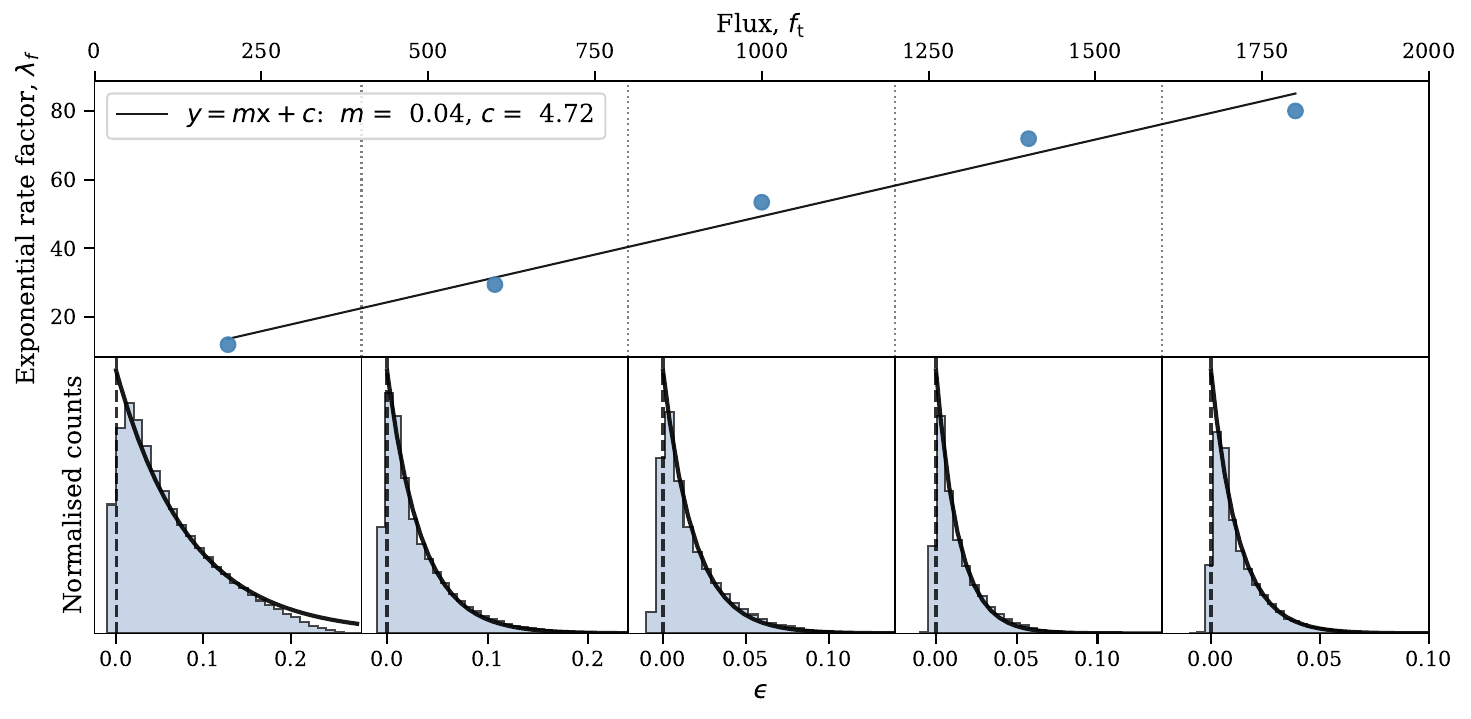}
    \caption{Flux evolution of the random exponential component to the noise, $\epsilon$. The lower panels show histograms of $\epsilon$ for flux bins of width 400. The upper panel shows the corresponding exponential scale factor, $\lambda_f$, for each flux bin, which is fit with a linear model, \mbox{$\lambda_f = m \cdot f_\mathrm{t} + c$}.}
    \label{figure:error_model_exp}
\end{figure*}

\begin{figure*}[htbp]
\centering
\includegraphics[width=0.85\textwidth]{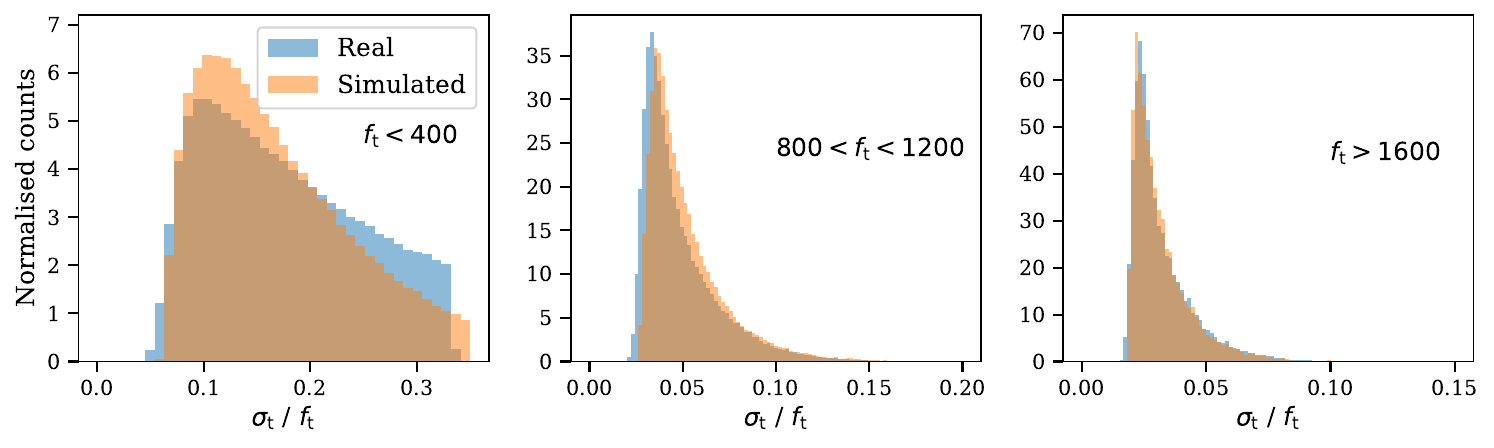}
    \caption{Comparison of the simulated and real $\sigma_\mathrm{t}/f_\mathrm{t}$ values at flux bins in the low ($f_\mathrm{t}<400$), medium ($800<f_\mathrm{t}<1200$), and high ($f_\mathrm{t}>1600$) flux regimes.}
    \label{figure:error_model_hist}
\end{figure*}

Due to the correlated nature of the error model in Equation~\ref{equation:errmodel}, it is not possible to determine the optimal values of the parameters $(e_\mathrm{b}, m, c, \delta)$ through minimisation or maximum-likelihood methods. Instead, we fit the underlying, deterministic model in Equation~\ref{equation:errmodel_basic} to the original ZTF light curves\footnote{Throughout this work, we use the term “flux” for consistency with common usage in optical astronomy, although our measurements are in flux densities. These are expressed in janskys and scaled to be consistent with the AB magnitude system.}, to recover the parameter $e_\mathrm{b}$. We then subtracted this model from the data to study the behaviour of the remaining noise. The left panels of Fig.~\ref{figure:error_model} show the distribution of $\sigma_\mathrm{t}/f_\mathrm{t}$ for the ZTF sample, with the residual distribution for $\epsilon$ shown in the lower panel. As illustrated in Fig.~\ref{figure:error_model_exp}, we fit an exponential curve to the $\epsilon$ component of the data in flux bins to determine the parameters $m$ and $c$. Finally, we compared the simulated data to the real data at different flux bins to determine the offset parameter $\delta$. This is illustrated in Fig.~\ref{figure:error_model_hist}, for binned data in the low ($f_\mathrm{t}<400$), medium ($800<f_\mathrm{t}<1200$), and high ($f_\mathrm{t}>1600$) flux value ranges. This process was repeated iteratively until the optimum parameters were derived, which are given as $e_\mathrm{b}=18.0$, $m=0.04$, $c = 4.7$, $\delta=-0.006$.

Figure~\ref{figure:error_model} shows a comparison of the $\sigma_\mathrm{t}/f_\mathrm{t}$ vs. $f_\mathrm{t}$ distributions of the original ZTF light curves and the ZTF light curves with flux errors simulated according to Equation~\ref{equation:errmodel}. We note that the errors are slightly underestimated at the lower flux range of $f_\mathrm{t}<400$, which can also be seen in the left panel of Fig.~\ref{figure:error_model_hist}. In particular, Fig.~\ref{figure:error_model_hist} shows that this primarily affects measurements with $\sigma_\mathrm{t}/f_\mathrm{t} > 0.2$, which corresponds to data points with a signal-to-noise ratio (SNR) of less than five. This is likely because the noise is not purely exponential within this range, as shown in the bottom left panel of Fig.~\ref{figure:error_model_exp}, where the exponential curve is a poorer fit in the $f_\mathrm{t}<400$ flux bin. This could be addressed by including an additional random Gaussian component dependent on $f_\mathrm{t}$ in the error model, but this was not implemented since the parameter optimisation would become too complex. Data points with low SNRs are effectively downweighted by the model during both training and inference, so they are unlikely to strongly influence the results.

Using our empirically-determined error model, we can simulate how the flux uncertainties would appear if an object were observed at a higher redshift. To approximate the corresponding flux values under these conditions, we apply a redshift-motivated scaling to the light curves. Specifically, we define:
\begin{equation}
\label{equation:D}
    f_\mathrm{z} = D \: f_\mathrm{t} , \quad \text{where} \quad D = \frac{D_L(z_{\rm true})^2}{D_L(z_{\rm sim})^2},
\end{equation}
with $D_L$ the luminosity distance. This is computed assuming a flat $\Lambda \mathrm{CDM}$ cosmology with parameters from \citet{Planck2020}: $H_0=67.7 \: \mathrm{km\,s^{-1}Mpc^{-1}}$ and $\Omega_M=0.310$. The $D_L^2$ factor accounts for luminosity-distance dimming, since the observed flux of a source with fixed intrinsic luminosity scales as \(F \propto D_L^{-2}\). We separately compute K-corrections to account for the shifting of the observed bandpass relative to the rest-frame spectral energy distribution (SED) using spectral templates (see Sect.~\ref{subsection:kcorr}), and apply these as an additive magnitude offset after flux scaling. 

While this formula does not attempt to model the full redshift transformation (including, e.g., time dilation of the observation dates), it provides an approximation of redshift evolution that captures the dominant dimming effect without altering the cadence of the light curve. This enables consistent training and evaluation of classification models with light curves at similar epochs but varying SNR. Additionally, the majority of transients in our sample lie at relatively low redshifts ($z \lesssim 0.2$), and our simulated redshift shifts are modest ($z_{\rm sim} - z_{\rm true} \leq 0.1$), corresponding to time dilation effects of less than $\sim$10\%. These effects are typically subdominant compared to photometric uncertainties and cadence limitations.

Combining Equations~\ref{equation:errmodel} and \ref{equation:D}, the scaling of the flux uncertainties when transforming a measured flux $f_\mathrm{t}$ at redshift $z_\mathrm{true}$ to the simulated flux $f_\mathrm{z}$ at $z_\mathrm{sim}$ is given by

\begin{equation}
    \label{equation:new_err}
    \frac{\sigma_\mathrm{z}}{\sigma_\mathrm{t}}= \frac{\sqrt{D\:f_\mathrm{t} + e_\mathrm{b}^2} + D\:f_\mathrm{t}\:(\epsilon + \delta)}{\sqrt{f_\mathrm{t} + e_\mathrm{b}^2} + f_\mathrm{t}\:(\epsilon + \delta)}.
\end{equation}
The new flux error, $\sigma_\mathrm{z}$, is obtained by multiplying the original flux error by the factor in Equation~\ref{equation:new_err}.

\begin{figure*}
\begin{center}
\resizebox{\hsize}{!}{\includegraphics{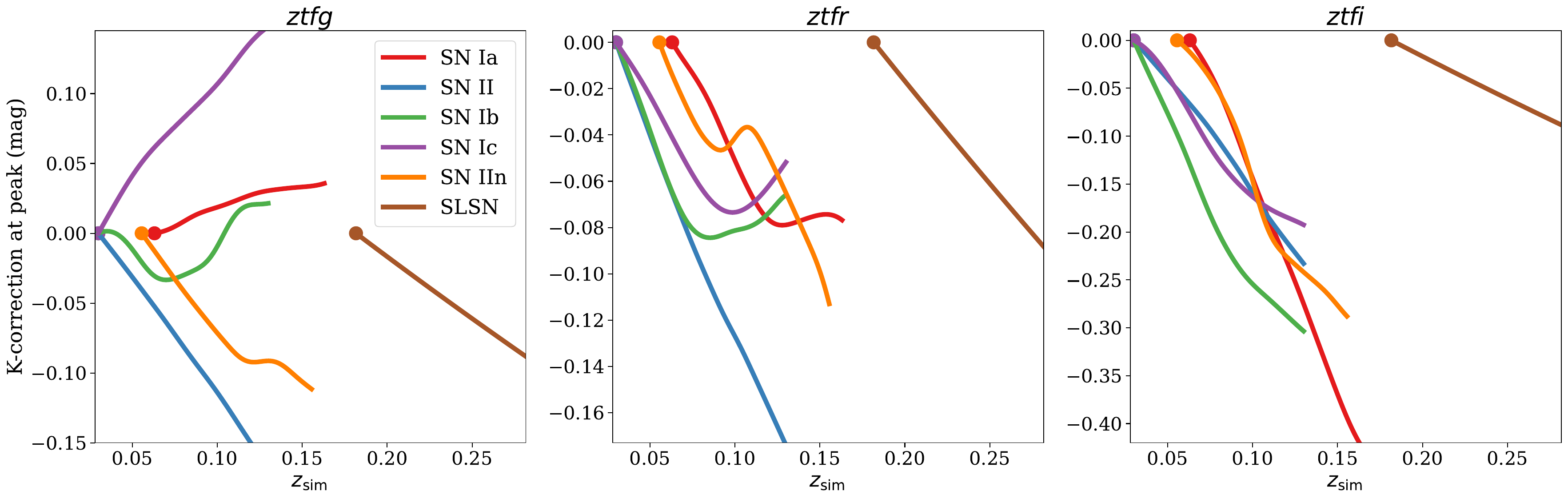}}
  \caption{K-corrections at peak brightness for the three ZTF bands ($g$: left, $r$: centre, $i$: right), applied to a representative set of SN types. For each class, the true redshift $z_\mathrm{true}$ (indicated by a dot at $(z_\mathrm{true}, 0)$) is taken as the median redshift from the BTS catalogue. Corrections are shown over the range $z_\mathrm{sim} \in [z_\mathrm{true}, z_\mathrm{true}+0.1]$. Template corrections are shown for SNe~Ia (red), II (blue), Ib (green), Ic (purple), and IIn (orange), alongside the constant approximation adopted for SLSNe (brown).
  }
     \label{figure:kcorr}
\end{center}
\end{figure*}

\subsection{K-correction}
\label{subsection:kcorr}

A K-correction is the conversion of flux measured in an observer-frame passband to flux in a rest-frame passband, accounting for the shift in wavelength at different redshifts \citep{Oke1968}. We applied a K-correction during redshift augmentation to account for the shift in rest-frame wavelength at higher simulated redshifts. This was performed using spectral templates from \texttt{sncosmo} \citep{sncosmo}\footnote{\href{https://sncosmo.readthedocs.io}{\texttt{sncosmo.readthedocs.io}}}, where available.

The source model was constructed using an \texttt{sncosmo} template, selected according to the SN type (see Table~\ref{table:appendix_kcorr} in the Appendix for details on the specific templates used). For the non-parametric models (i.e., all except the SALT2 model used for SNe~Ia), we did not need to choose specific parameter values, since we compute a relative quantity and the model amplitude cancels in the K-correction. For the SALT2 model \citep{Guy2007,Betoule2014}, we fixed the parameters to $[x_1, c, M_B]=[1.0,0.2,-19.4]$. While individual SNe within each class can exhibit spectral diversity (particularly in colour for core-collapse SNe, and in both light curve shape and colour for SNe~Ia), these templates provide a reasonable approximation for capturing average K-correction behaviour across the population. K-corrections were determined for each object by computing the difference between the model-predicted observed flux at the original redshift ($z_\mathrm{true}$) and the simulated redshift ($z_\mathrm{sim}$).

There are no templates available for SLSNe in \texttt{sncosmo}, due to their diverse and variable spectral properties. Instead, we relied upon an approximation of the K-correction formulae described in \citet{Hogg2002}, where we would expect a constant correction of $-2.5 \: \log{(1+z)} \, \mathrm{mag}$. Previously, \citet{Chen2023} showed that this approximation is comparable to K-corrections computed from spectra for SLSNe-I observed by ZTF. Therefore, we applied a correction of
\begin{equation}
    K = -2.5 \: \log{\left(\frac{1 + z_\mathrm{sim}}{1 + z_\mathrm{true}}\right)} 
\end{equation}
when we transform our SLSN light curves from $z_\mathrm{true}$ to $z_\mathrm{sim}$.

Figure~\ref{figure:kcorr} illustrates the K-corrections at peak brightness for the three ZTF bands, applied to a representative set of SN types. For each class, the true redshift $z_\mathrm{true}$ is taken as the median redshift in the BTS catalogue, and corrections are shown over the range $[z_\mathrm{true}, z_\mathrm{true} + 0.1]$ in simulated redshift $z_\mathrm{sim}$. The approximation used for SLSNe is justified, since the K-corrections across different SN types are broadly similar in magnitude. We also note that the $i$-band exhibits the largest corrections, though this is unlikely to significantly affect our results due to the relatively sparse $i$-band sampling.

\subsection{Noisification}
\label{subsection:noisify}

In this subsection, we outline the method for adding noise to the light curves, a technique referred to as `noisification'. This involves selecting specific augmentation parameters to generate different versions of the training data. The evaluation of these parameter choices is presented in Sect.~\ref{subsection:results_param_opt}, where we selected the optimal augmented datasets for our final models.

First, we selected a new redshift value by sampling within the range $z_\mathrm{sim} \in [z_\mathrm{true}, z_\mathrm{true} + 0.1]$, using a probability distribution proportional to $z^\mathrm{\texttt{z\_scale}}$, where $\mathrm{\texttt{z\_scale}}\in\{0,2\}$. We tested two distributions: for $\mathrm{\texttt{z\_scale}}=0$, the distribution is flat, implying no preference across the interval; for $\mathrm{\texttt{z\_scale}}=2$, the distribution reflects the expected scaling of SNe in the low-redshift universe ($z \lesssim 0.3$), assuming a constant volumetric SN rate. In this regime, the number of observed SNe scales with the comoving volume element, $dV \propto z^2 \, dz$. The choice of $\mathrm{\texttt{z\_scale}}=2$ preferentially samples higher redshifts within the interval.

Next, we generated a simulated light curve from the new redshift value. We calculated the new flux error, $\sigma_\mathrm{z}$, by multiplying the observed flux error by the scale factor in Equation~\ref{equation:new_err}, as described in Sect.~\ref{subsection:errmodel}. The new flux value, $f_\mathrm{z}$, is the sum of the scaled flux and a Gaussian scatter term, given by $f_\mathrm{z}=D\:f_\mathrm{t} + \mathcal{N}\:(0, \sigma_\mathrm{z}^2)$, where $D$ is defined in Equation~\ref{equation:D}. This procedure adds simulated noise on top of the original measurement uncertainties, resulting in slightly higher total scatter than a real observation at the simulated redshift. However, since the original noise cannot be disentangled from the observed flux, removing it is not feasible. This conservative approach ensures we do not underestimate uncertainties when simulating higher-redshift observations. A K-correction term was also applied, as described in Sect.~\ref{subsection:kcorr}.

We applied quality cuts to the simulated light curves, requiring the peak detection to have an SNR greater than five. Furthermore, each light curve was required to contain at least five detections with an SNR greater than five. Simulated light curves that did not meet these criteria were excluded from the training sample.

To ensure that the simulated light curves remain diverse, even when generating hundreds of copies, we applied additional augmentation by removing data points. This was done using two approaches: based on local density or via random subsampling. For the density-based method, we computed the number of detections per band within a rolling five-day window. If the local density within a window exceeded a user-defined threshold, called \texttt{cadence\_scale}, data points were dropped at random with a uniform probability. We tested two values for \texttt{cadence\_scale}: 0.5 (corresponding to one detection every two days per band) and 100 (which results in no density-based dropping). For random subsampling, the parameter \texttt{subsampling\_rate} specifies the fraction of data points retained. We only removed data points randomly if there were more than 10 detections in any band. For the \texttt{subsampling\_rate}, we tested values of 0.7, 0.9, and 1.0 (i.e., no subsampling).

Figures~\ref{figure:example_Ia} and \ref{figure:example_II} in the Appendix present examples of an augmented SN~Ia and SN~II, respectively. For comparison, each panel also includes a real light curve at a similar redshift, illustrating that the augmented light curves exhibit realistic brightness evolution and scatter consistent with real observations.

\subsection{Testing the noisification with SALT2 fits to SNe~Ia}
\label{subsection:salt2fits}

As an additional test of our error model, we compared SALT2 fits to the original SN~Ia sample and an augmented version of the same sample. The SALT2 model \citep{Guy2007} is a light curve template that characterises SNe~Ia using two parameters: stretch ($x_1$) and colour ($c$). In addition to these, the model includes a normalisation factor $x_0$, which determines the overall brightness of the SN, and a time of peak brightness $t_0$. We performed the fits using \texttt{sncosmo} \citep{sncosmo}, implementing SALT2 version 2.4 as trained by \citet{Betoule2014}.

We followed the procedure described in Sect. \ref{subsection:noisify} to generate an augmented copy of each SN~Ia using the following augmentation parameters: $\mathrm{\texttt{z\_scale}}=2,\: \mathrm{\texttt{cadence\_scale}}=100,\: \mathrm{\texttt{subsampling\_rate}}=1$ (i.e., we did not manually remove data points). For each fit, we defined the initial parameters for the SALT2 fit as follows: $t_0=t_{0,\mathrm{est}}$, $x_0=1.0$, $x_1=0.0$, $c=0.0$, where $t_{0,\mathrm{est}}$ is estimated as the time of the brightest detection in any filter. We also restricted the fitting window to the rest-frame phase range $t\in[-20, 50]$, which corresponds to the model limit of SALT2. Fits were performed on both the original and augmented light curves, and the corresponding reduced $\chi^2$ values were calculated.

\begin{figure}[t!]
\centering
\includegraphics[width=\hsize]{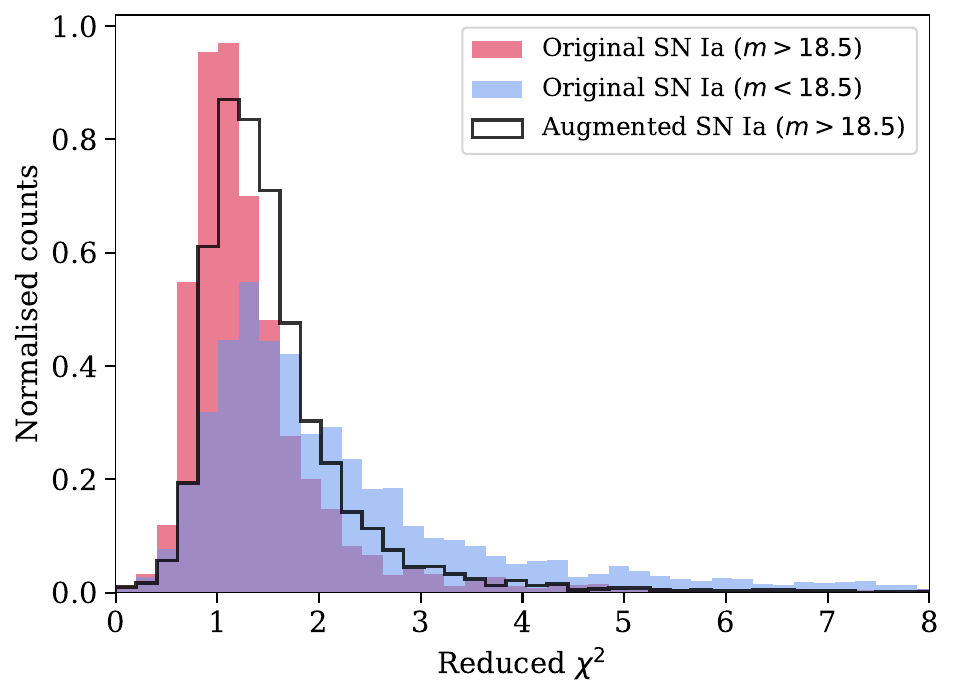}
    \caption{Comparison of the reduced $\chi^2$ values from SALT2 fits to the original ZTF SN~Ia sample -- divided into bright (blue) and faint (red) subsamples -- and to the augmented copies of the same sample (black outline).}
    \label{figure:salt2_chi}
\end{figure}

A comparison of the reduced $\chi^2$ values from the SALT2 fits to the original and augmented samples is shown in Fig. \ref{figure:salt2_chi}, where we further divide the ZTF sample into bright ($m<18.5$) and faint ($m>18.5$) subsamples. This division highlights differences in the reduced $\chi^2$ distributions: the brighter SNe~Ia tend to exhibit larger values and a broader overall distribution. This is likely due to calibration issues, such as detector nonlinearity, which introduce systematic errors. These systematics have a greater impact on the light curves of brighter objects because of their higher SNRs, whereas fainter, noise-dominated light curves are less affected by these systematics due to their larger observational uncertainties. We also show the augmented sample, limited to objects with $m>18.5$.

We note that the distribution of reduced $\chi^2$ values for the augmented sample is in reasonable agreement with that of the faint ZTF subsample. However, the augmented sample has a slightly higher median reduced $\chi^2$ (1.39 compared to 1.18), likely due to underestimated errors at low fluxes, as discussed in Sect.~\ref{subsection:errmodel}. Additionally, as discussed in Sect.~\ref{subsection:noisify}, the augmented light curves include both the original measurement noise and additional simulated scatter, which may slightly overestimate the total scatter. It is also possible that this discrepancy arises from the broader and fainter magnitude range of the augmented SNe~Ia. As the SNR decreases near the ZTF detection limit, the number of significant detections, and thus the degrees of freedom in the SALT2 fit, tends to decrease, which can inflate the reduced $\chi^2$. Overall, these results suggest that our noisification method produces realistic ZTF light curves at higher redshifts.

\section{Classification procedure}
\label{section:classification_method}
In this section, we describe the generation of the training and test samples. We then detail how the training sample was used to train a \texttt{ParSNIP} model that captures transient behaviour. Finally, we explain how features extracted from the trained \texttt{ParSNIP} model were used to train a classifier and classify the test sample.

\subsection{Generating the training and test samples}
\label{subsection:create_train}
First, we divided the data into training and test samples, as described in Sect.~\ref{subsection:traintest}. This split is performed before any augmentation to ensure that no copies of the same object appear in both sets.

We next determined the number of augmented copies to generate per ZTF object, based on its class. We tested three augmentation strategies: (1) balanced, where the number of objects in each class is equalised by generating more copies of underrepresented classes (e.g., significantly more SLSNe than SNe~Ia); (2) unbalanced, where a constant number of augmented copies is created per class, preserving the original class imbalance; and (3) hybrid, where non-Ia classes are balanced, but approximately five times more SNe~Ia are generated, with the goal of improving the classifier’s ability to distinguish them from other types.

Before augmentation, we imposed phase limits based on the peak date from the BTS catalogue (defined as the date of the brightest detection in any filter), to exclude FP data well outside the transient event, such as remaining pre- or post-explosion baselines or spurious detections. For longer-lived classes such as SLSNe and SNe~IIn, we used a phase range of $t\in[-100, 200]$ days relative to peak; for all other classes, the range was $t\in[-50, 80]$ days. While these cuts do not explicitly provide the classifier with information about the transient’s true duration or peak date, the longer phase range for SLSNe and SNe IIn means that their light curves typically include data over a longer timespan. Consequently, the classifier could indirectly infer the longer duration characteristic of these classes based on the extent of the observed light curve. However, this information is a physically meaningful feature of the transient rather than an explicit input, and the model still relies on the photometric data itself to distinguish classes.

\begin{figure}[t!]
\centering
\includegraphics[width=\hsize]{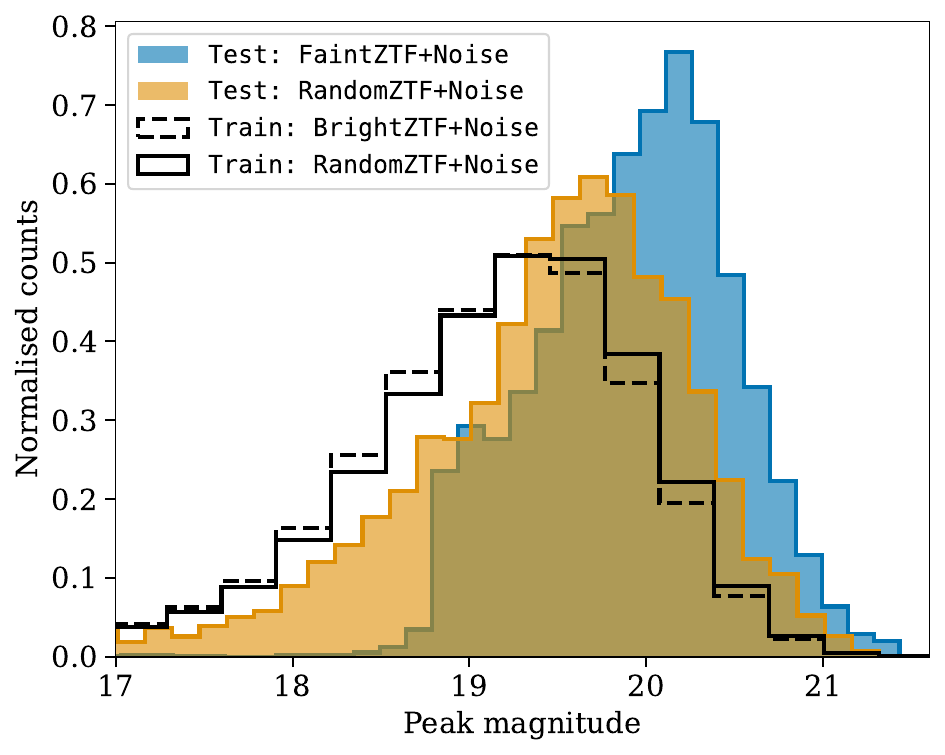}
    \caption{Comparison of training and test peak apparent magnitude distributions between the \texttt{BrightZTF}/\texttt{FaintZTF} and a representative \texttt{RandomZTF} split (one random seed shown). The training samples were augmented using the parameters shown in bold in Table~\ref{table:aug_params}.}
    \label{figure:hist_traintest}
\end{figure}

Each training light curve was augmented as described in Sect.~\ref{subsection:noisify}. We aimed to generate the required number of augmented light curves, but this was sometimes not possible for objects that were already faint. If a light curve passing the SNR cuts was not produced within the first 50 iterations or during any subsequent 2000 consecutive iterations, we stopped the process and moved on to the next object.

We applied the same augmentation methodology to the test light curves but fixed the augmentation parameters to ensure each test sample was identical, allowing direct comparison of model performance. We used the following augmentation parameters, as in Sect.~\ref{subsection:salt2fits}: $\mathrm{\texttt{z\_scale}}=2,\: \mathrm{\texttt{cadence\_scale}}=100,\: \mathrm{\texttt{subsampling\_rate}}=1$. For class scaling, we used an unbalanced dataset with 10 copies per class to preserve the original class imbalance, reflecting a more realistic distribution. We also chose $\mathrm{\texttt{z\_scale}}=2$ to better reflect the expected redshift distribution.

Furthermore, we imposed additional quality cuts on the original test data prior to augmentation, requiring more than five detections with an SNR of greater than five, a duration (defined as the time span between its first and last detection) exceeding 20 days, and observations in at least two bands, each with a minimum of two detections. These cuts were motivated by the requirement for light curves suitable for cosmological analyses.

Figure~\ref{figure:hist_traintest} shows the peak apparent magnitude distributions for the different training and test configurations, highlighting the fainter magnitude range in the \texttt{FaintZTF+Noise} test sample. While the \texttt{BrightZTF+Noise} training set is slightly skewed toward brighter magnitudes, its distribution is broadly similar to that of the \texttt{RandomZTF+Noise} training set.

\subsection{Training the \texttt{ParSNIP} model and classifier}
\label{subsection:parsnip}
\begin{figure}[t!]
\centering
\includegraphics[width=\hsize]{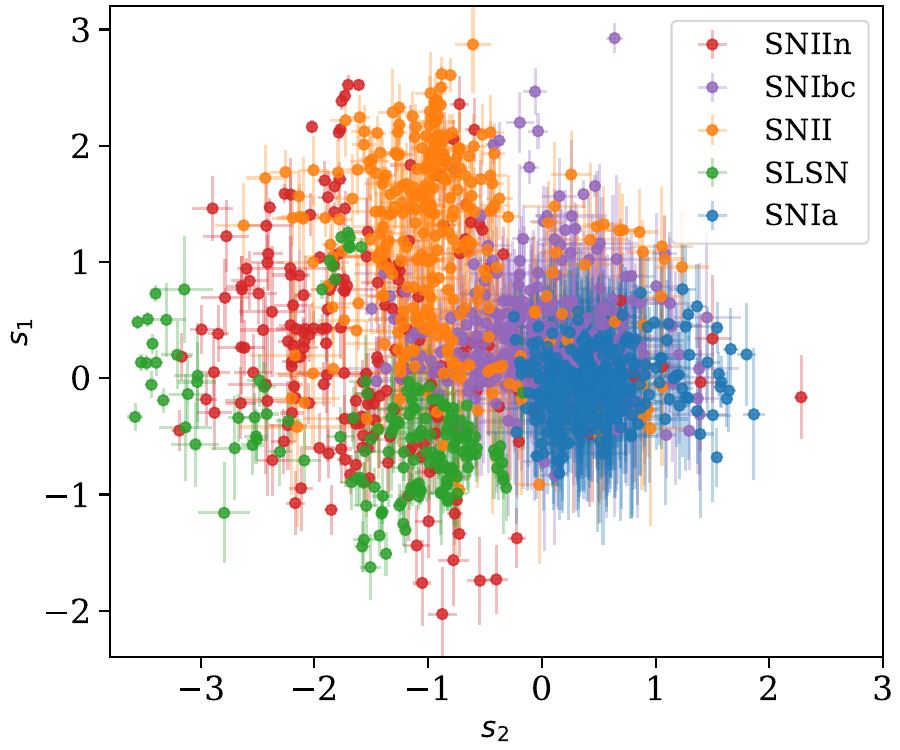}
    \caption{Distribution of transient classes in the intrinsic latent space defined by the $s_1$ and $s_2$ parameters of the trained \texttt{ParSNIP} model. Each point represents a light curve from the augmented \texttt{FaintZTF+Noise} test sample, coloured by class: SN~Ia (blue), SN~II (orange), SN~Ib/c (purple), SLSN (green) and SN~IIn (red). The separation between classes emphasises the model’s ability to distinguish between transient types, though some overlap remains (e.g., between SLSNe, SNe~II, and SNe~IIn, or between SNe~Ia and SNe~Ibc).}
    \label{figure:latent_res}
\end{figure}

We trained \texttt{ParSNIP} models using our training datasets, following a methodology similar to \citet{Boone2021}. \texttt{ParSNIP} is a modified version of a variational autoencoder \citep{Kingma2013} that is designed to capture the diverse behaviour of transients using a low-dimensional latent space. Its architecture includes a neural network that encodes each light curve into three intrinsic latent parameters, referred to as $s_1$, $s_2$, and $s_3$, which describe the intrinsic spectrum of a given transient. There is an additional component, denoted the `physics layer', which models how the light from the transient would be observed, by incorporating explicit latent variables (a reference time $t_0$, amplitude of the light curve $A$, and the colour $c$) along with observational metadata (redshift $z$, observation time $t$, and bandpass $B$). The physics layer enables \texttt{ParSNIP} to learn a representation of each transient that is invariant to observing conditions, such as redshift. Figure~\ref{figure:latent_res} illustrates how two of the intrinsic latent parameters, $s_1$ and $s_2$, vary across the different classes in our model.

To train the model, we adopted the hyperparameters from \citet{Boone2021}, who found that model performance is generally insensitive to these settings. We used the Adam optimiser \citep{Kingma2014} with an initial learning rate of $10^{-3}$ and a batch size of 128 light curves. The learning rate was reduced by a factor of 0.5 if the loss function did not improve after 10 epochs, continuing until the learning rate dropped below $10^{-5}$. Consistent with \citet{Boone2021}, we found that models typically converged after approximately 300 epochs, although this varied considerably across training runs.

After training the model, we extracted the \texttt{ParSNIP} features for both the training and test samples. These features were then used to train a classifier using the gradient-boosted decision tree framework \texttt{LightGBM} \citep{Ke2014}. Following the approach of \citet{Boone2019}, we adopted the default hyperparameters, employed 10-fold cross-validation, and set the \texttt{min\_child\_weight} parameter to 100 to prevent any overfitting to the augmented light curves. The classification features include colour, luminosity, amplitude, the three latent variables, their associated uncertainties, and the uncertainty in the reference time. \texttt{ParSNIP} also outputs the total number of detections, along with counts of detections with SNR > 3 and > 5 across three epochs: 50 days before peak, 50 days after peak, and all following days. These values were included as additional classifier features.

Our primary classifier was a binary model distinguishing SNe~Ia from all other types, outputting the probability that a given object is an SN~Ia. We also tested multi-class classifiers, which assign probabilities that an object belongs to each class in the training set: SN~Ia, SN~Ib/c, SN~II, SN~IIn, and SLSN.

\section{Results and discussion}
\label{section:results}
In this section, we present the results of our classification procedure. We begin by describing the optimisation of the augmentation parameters (Sect.~\ref{subsection:results_param_opt}) and show the results from one of our best-performing models (Sect.~\ref{subsection:results_best}). We then evaluate the classifier on additional datasets, including the ZTF SN~Ia DR2 sample (Sect.~\ref{subsection:results_dr2}) and the peculiar or uncommon subtypes from BTS that we initially excluded (Sect.~\ref{subsection:results_contam_pec}). Finally, we discuss potential applications of our method in supernova cosmology (Sect.~\ref{subsection:results_cosmo}) and real-time survey classification (Sect.~\ref{subsection:results_livetest}).

The primary metrics used to evaluate classifier performance are recall (also referred to as efficiency or completeness, the fraction of the real events we correctly identify) and precision (also referred to as purity, the fraction of predicted positive classifications that are correct). These are defined as:
\begin{equation}
    \mathrm{recall}=\frac{\mathrm{TP}}{\mathrm{TP}+\mathrm{FN}}, \:\:\:\:\: \mathrm{precision}=\frac{\mathrm{TP}}{\mathrm{TP}+\mathrm{FP}},
\end{equation}
where TP is the number of true positives, FN is the number of false negatives, and FP is the number of false positives. Throughout multiple iterations, we observed a trade-off between SN~Ia recall and the recall of other classes -- improvements in one often led to declines in the other. Given the context of our work, we prioritised SN~Ia recall, as discussed in Sect.~\ref{subsection:intro_snia}, since this metric has a more direct impact on the effectiveness of SN~Ia cosmology analyses.

As described in Sect.~\ref{subsection:traintest}, we trained and tested on different samples to evaluate the robustness of our models. We followed two main strategies: (1) training on \texttt{BrightZTF} or its augmented counterpart, \texttt{BrightZTF+Noise}, and testing on \texttt{FaintZTF} or \texttt{FaintZTF+Noise}; and (2) training and testing on a 90\%/10\% split of \texttt{RandomZTF+Noise}.

\subsection{Augmentation parameter optimisation}
\label{subsection:results_param_opt}
We tested various augmentation parameters (described in Sects.~\ref{subsection:noisify} and \ref{subsection:create_train}) to identify which values produced the best classification performance. The tested parameters and their values are listed in Table~\ref{table:aug_params}. We evaluated model performance using the \texttt{FaintZTF} and \texttt{FaintZTF+Noise} test sets, focusing exclusively on results from the binary SN~Ia classifier. The primary metrics were the mean SN~Ia recall and the mean SN~Other recall, averaged over both test sets.

\begin{table}[h]
\caption{\label{table:aug_params} Summary of augmentation parameter configurations tested in this study.}
\centering
\begin{tabular}{c | c  c}
\hline\hline
Parameter & &Values \\
\hline
Class scale & \texttt{Balanced}: & 2, 3, 4, 5 \\
 & \texttt{Unbalanced}: & 2, 5, 10, 15 \\
 & \texttt{Hybrid}: & \textbf{10} \\
\texttt{z\_scale} && \textbf{0}, 2 \\
\texttt{cadence\_scale} && 0.5, \textbf{100} \\
\texttt{subsampling\_rate} && 0.7, \textbf{0.9}, 1 \\
\hline
\end{tabular}
\tablefoot{The bolded values indicate the parameters of the best models discussed in Sect.~\ref{subsection:results_best}.}
\end{table}

For balanced class scaling, the number of SN~Ia augmentations is set by the scaling factor (e.g., a scale of 3 produces three versions per SN~Ia), and all other classes are upsampled to match this total, resulting in equal numbers of objects per class. In the unbalanced setting, the same number of augmentations is applied to each object across all classes (e.g., scale = 5 produces five versions per object, regardless of class). The hybrid strategy generates 10 versions per SN~Ia, while the remaining classes are augmented to approximately match a sample in which each class contains twice the original number of SN~Ia objects. In other words, the hybrid scaling generates five times as many SN~Ia as each of the other individual classes.

\begin{figure}[t!]
\centering
\includegraphics[width=\hsize]{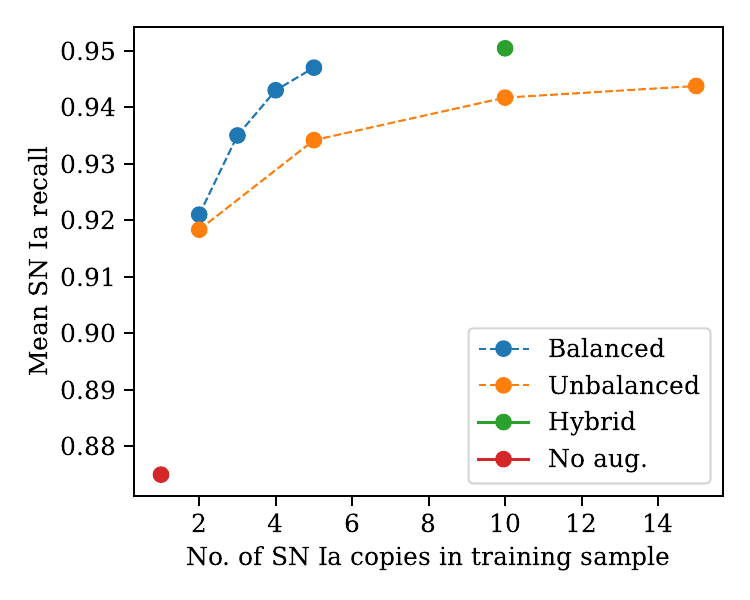}
  \caption{Mean SN~Ia recall (averaged over multiple iterations of the \texttt{FaintZTF} and \texttt{FaintZTF+Noise} test sets) as a function of the number of SNe~Ia we generate in the \texttt{BrightZTF+Noise} training sample. Results are shown for the different class scaling strategies described in the main text: balanced (blue), unbalanced (orange), and hybrid (green). For comparison, we also include a model trained without augmentation (i.e., on \texttt{BrightZTF} only), denoted `no aug.' (red), to illustrate the benefit of augmentation.
      }
 \label{figure:classvsrecall}
 \end{figure}

We found that class scaling had the greatest impact on classification performance, highlighting the effectiveness of augmentation in improving the classifier. As shown in Fig.~\ref{figure:classvsrecall}, the mean SN~Ia recall increases with the number of SNe~Ia generated. Among the strategies tested, the hybrid approach achieved the best average performance, followed by the balanced class scaling. Conversely, the unbalanced class scaling produced the highest average SN~Other recall.

We also observed a weaker dependence on the \texttt{cadence\_scale} and \texttt{subsampling\_rate} parameters. This is illustrated for the balanced classes in Figs.~\ref{figure:cadencevssubsamp_Ia} and \ref{figure:cadencevssubsamp_nonIa} in the Appendix, where the average SN~Ia recall was highest when no data points were removed, while the average SN~Other recall improved with more aggressive subsampling.

Introducing scatter in the observation times of the detections was also tested. However, even small perturbations (e.g., 0.1 days, or approximately 2.4 hours) degraded the SN~Ia precision. This effect likely arises because the temporal offsets shift the observed fluxes in different bands relative to each other, altering the inferred colours and the shape of the light curves. Since classification relies on subtle differences in colour evolution and light curve features, these shifts can increase confusion with SNe~Ib/c.

\subsection{Classification results}
\label{subsection:results_best}

After evaluating different augmentation parameters in Sect.~\ref{subsection:results_param_opt}, we selected the model configuration corresponding to the bolded parameters in Table~\ref{table:aug_params}. To demonstrate the robustness and generalisability of our classification model, we present four scenarios:
\begin{enumerate}[(i)]
    \item training and testing on the original, unaugmented data (Train: \texttt{BrightZTF}, Test: \texttt{FaintZTF}),
    \item training on original data but testing on augmented (noisier) light curves (Train: \texttt{BrightZTF}, Test: \texttt{FaintZTF+Noise}),
    \item training on augmented data but testing on the real faint data (Train: \texttt{BrightZTF+Noise}, Test: \texttt{FaintZTF}),
    \item training and testing on augmented data (Train: \texttt{BrightZTF+Noise}, Test: \texttt{FaintZTF+Noise}).
\end{enumerate}

\begin{figure}[t!]
\centering
\includegraphics[width=0.9\hsize]{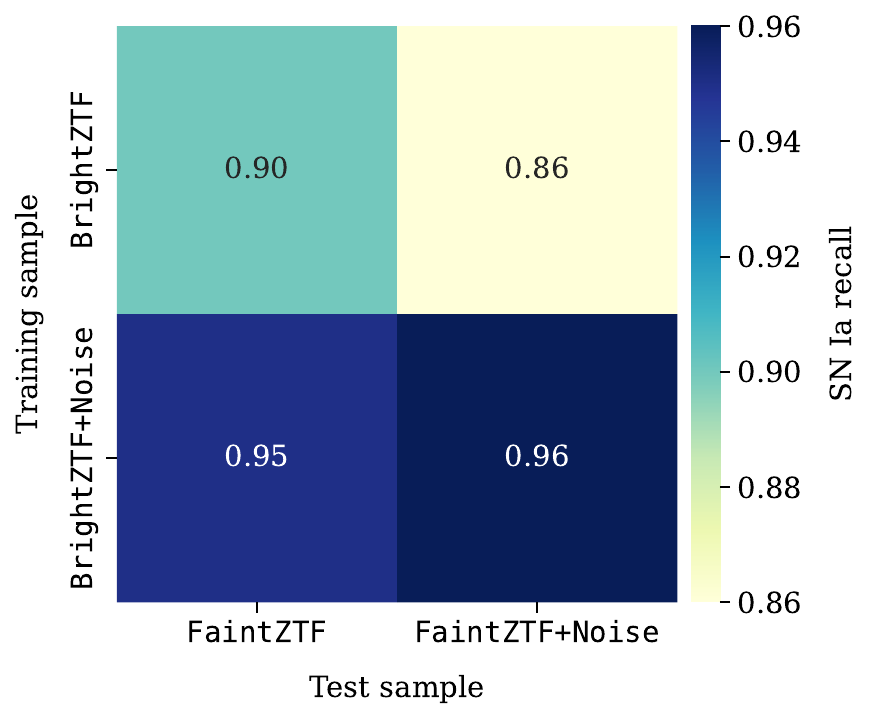}
  \caption{SN~Ia recall for our selected model configuration (bolded in Table~\ref{table:aug_params}) across different augmentation combinations of \texttt{BrightZTF} and \texttt{FaintZTF}. The four cases correspond to: (i) training and testing on unaugmented data, (ii) training on unaugmented data and testing on augmented data, (iii) training on augmented data and testing on unaugmented data, and (iv) training and testing on augmented data. The results demonstrate that noise augmentation in the training set improves classification performance on fainter, noisier test samples.
      }
 \label{figure:brightfaint_iarecall}
 \end{figure}

This framework allows us to isolate the effects of noise augmentation on both the training and test sets. Case (ii) evaluates how well a model trained on the real ZTF data performs on noisier data, highlighting its limitations in generalising to fainter flux regimes. Case (iii) examines how augmentation during training improves the model’s ability to classify real faint data. Finally, case (iv) reflects the most realistic use case, as the augmented test data are designed to more closely resemble the ZTF archive of unclassified transients. Figure~\ref{figure:brightfaint_iarecall} illustrates these results using the SN~Ia recall metric, demonstrating that our model accurately identifies SNe~Ia even with fainter, noisier light curves, and that noise augmentation in the training sample significantly improves performance. The SN~Ia recall increases from 90\% to 95\% on real data, and from 86\% to 96\% when tested on fainter, noise-augmented data. The corresponding confusion matrices across different augmentation combinations of \texttt{BrightZTF} and \texttt{FaintZTF} are presented in Fig.~\ref{figure:brightfaint_cm} in the Appendix. These results indicate a recall of 96\% and a precision of 99\% when training on \texttt{BrightZTF+Noise} and testing on \texttt{FaintZTF+Noise}.

\begin{figure*}[htbp]
    \centering
    \begin{subfigure}[b]{0.44\textwidth}
        \centering
        \caption{Train: \texttt{RandomZTF+Noise}, Test: \texttt{RandomZTF+Noise},\\ SN~Ia binary classifier normalised by true class totals.}
        \includegraphics[width=\linewidth]{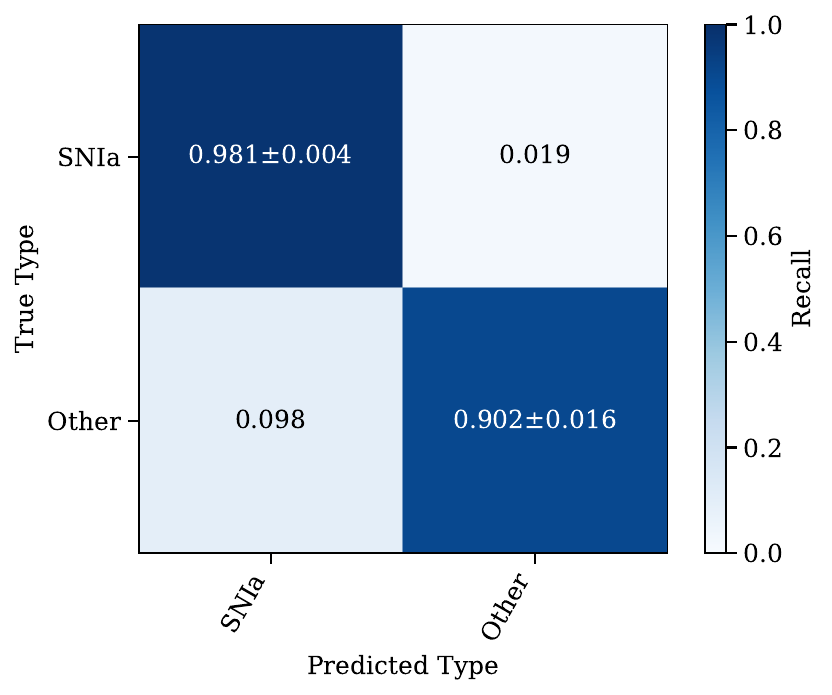}
    \end{subfigure}
    \hfill
    \begin{subfigure}[b]{0.44\textwidth}
        \centering
        \caption{Train: \texttt{RandomZTF+Noise}, Test: \texttt{RandomZTF+Noise},\\ SN~Ia binary classifier normalised by predicted class totals.}
        \includegraphics[width=\linewidth]{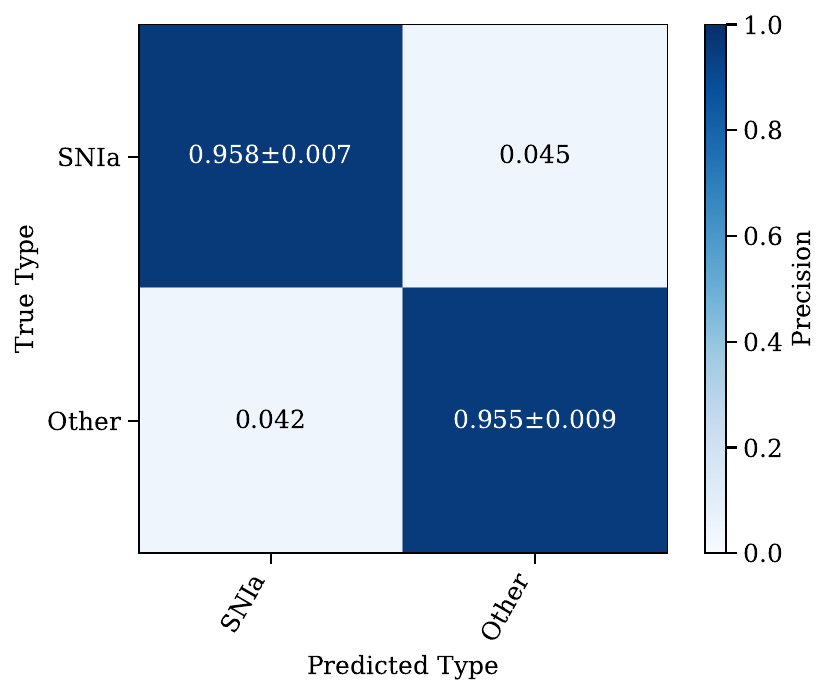}
    \end{subfigure}
    
    \vspace{0.1em}
    
    \begin{subfigure}[b]{0.46\textwidth}
        \centering
        \caption{Train: \texttt{RandomZTF+Noise}, Test: \texttt{RandomZTF+Noise},\\ Multi-class classifier normalised by true class totals.}
        \includegraphics[width=\linewidth]{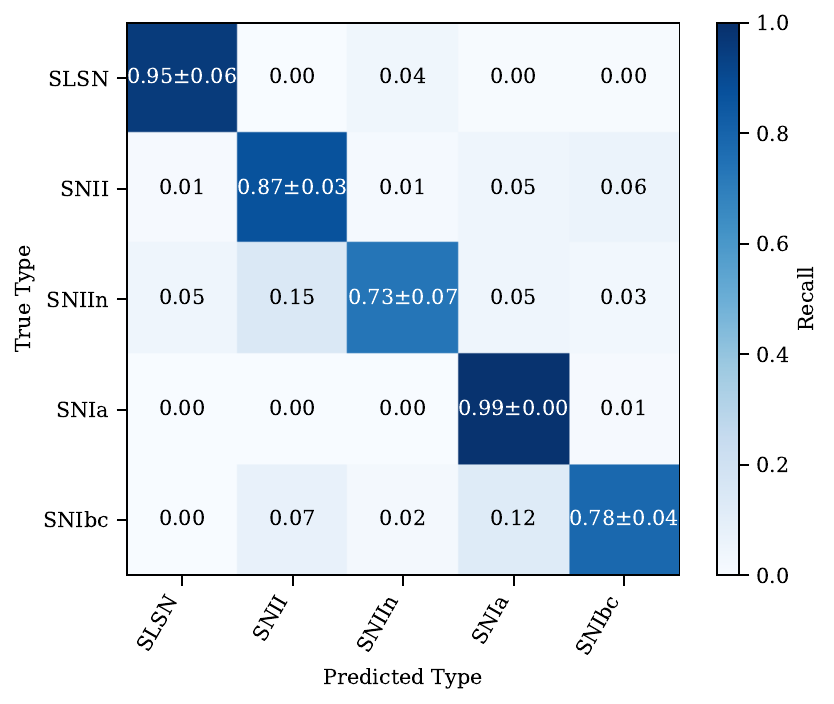}
    \end{subfigure}
    \hfill
    \begin{subfigure}[b]{0.46\textwidth}
        \centering
        \caption{Train: \texttt{RandomZTF+Noise}, Test: \texttt{RandomZTF+Noise},\\ Multi-class classifier normalised by predicted class totals.}
        \includegraphics[width=\linewidth]{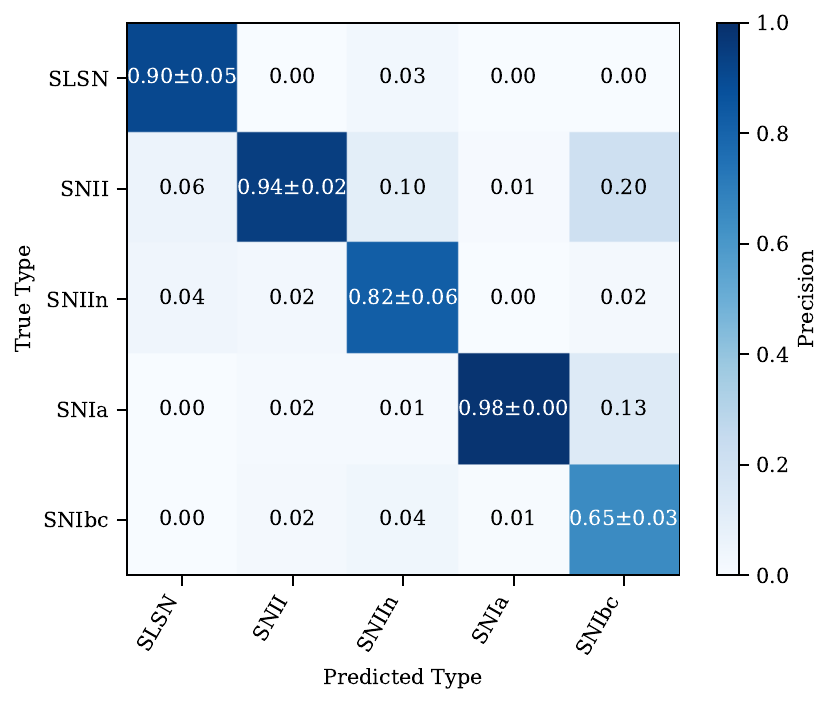}
    \end{subfigure}

    \caption{Confusion matrices averaged over five random train-test sets for both the SN~Ia binary (panels a and b) and multi-class classifiers (panels c and d). The matrices are normalised by true class totals to show the mean recall and its standard deviation (panels a and c), and by predicted class totals to show the mean precision and its standard deviation (panels b and d). }
    \label{figure:random_cm_all}
\end{figure*}

To ensure a robust evaluation across the full brightness range, we also performed multiple random splits (\texttt{RandomZTF}) of the data using five different seeds. This approach increases statistical confidence in the reported classification scores by averaging results over different train-test partitions, providing a more reliable estimate of the model’s overall performance. Figure~\ref{figure:random_cm_all} shows the confusion matrices for this configuration, for both the SN~Ia binary (panels a and b) and multi-class classifiers (panels c and d). The matrices are normalised by true class totals to show the mean recall and its standard deviation (panels a and c), and by predicted class totals to show the mean precision and its standard deviation (panels b and d).

For the SN~Ia binary classifier, we note that the SN~Ia recall is slightly improved at $(98.1 \pm 0.4\%)$. However, the SN~Other recall is slightly lower, with $(90.2\pm 1.6)\%$ vs. 97\% for the \texttt{BrightZTF+Noise} training with \texttt{FaintZTF+Noise} testing. This reduced SN~Other recall can be attributed to both the increased difficulty of learning from noisier, more heterogeneous training data and the intrinsic complexity of the SN~Other class. The \texttt{BrightZTF+Noise} training sample includes only the brightest 90\% of the light curves, which typically have higher SNR photometry. In contrast, the \texttt{RandomZTF+Noise} training set spans a wider brightness range prior to augmentation, including fainter examples of already diverse SN~Other classes, with lower SNR and sparser temporal sampling. This can introduce additional noise in the feature space, making it harder for the classifier to learn clear decision boundaries between SN~Other types and SNe~Ia, causing a decrease in the SN~Other recall.

In addition to differences in training data quality, the class composition of the SN~Other test sample also varies slightly between the two configurations. The \texttt{RandomZTF+Noise} test sets contain a higher fraction of SNe~Ibc (17.8\% vs.\ 14.5\%) -- a subclass more easily confused with SNe~Ia -- and a lower fraction of SNe~II (62.4\% vs.\ 66.0\%), which are generally easier to distinguish. This shift in subclass distribution likely contributes further to the reduced SN~Other recall in the \texttt{RandomZTF+Noise} configuration, as a larger fraction of the test set consists of more ambiguous examples.

For the multi-class classifier, we applied a minimum probability threshold of 0.7 to assign a classification. As a result, 10\% of objects remain unclassified. However, this threshold improves both recall and precision across all classes. With this configuration, 95.2\% of objects are correctly classified on average across five random seeds. This is comparable with the SN~Ia binary classifier, where 95.7\% of objects are correctly classified on average.

We can see from the multi-class classifier that the majority of contamination in the SN~Ia classification arises from the SN~Ib/c class. This is a well-known challenge in photometric classification due to the intrinsic similarities in the light curve shapes and colours of SNe~Ia and SNe~Ib/c, particularly when observed with limited or noisy data. Both classes can exhibit similar peak magnitudes and decline rates, making them difficult to distinguish solely from photometric features. Such misclassification can significantly contaminate SN~Ia cosmological samples and bias distance estimates. Reducing this contamination might require enhanced feature extraction, such as exploiting the second bump in redder bands, which is unique to SNe~Ia. We plan to explore this in future work.

The SN~Ia precision is $(95.8 \pm 0.7)\%$ from the SN~Ia binary classifier and $(98.1 \pm 0.4)\%$ from the multi-class classifier. While these values are encouraging, they are highly sensitive to the class fractions in the test sample. A more realistic estimate would require simulating the ZTF survey to determine the expected observed distribution of SNe. We leave this for future work.

 \begin{figure*}[htbp]
    \centering
    \begin{subfigure}[b]{0.47\textwidth}
        \centering
        \includegraphics[width=\linewidth]{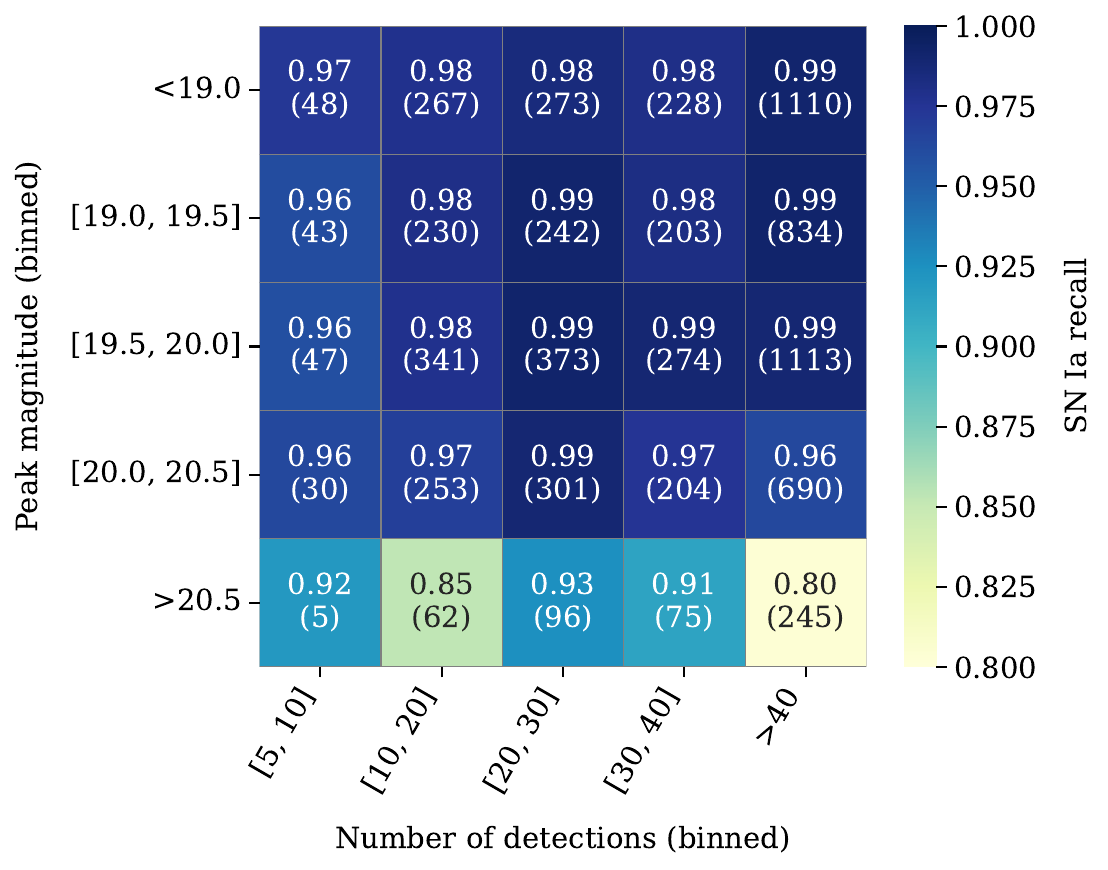}
        \caption{SN~Ia recall}
    \end{subfigure}
    \hfill
    \begin{subfigure}[b]{0.47\textwidth}
        \centering
        \includegraphics[width=\linewidth]{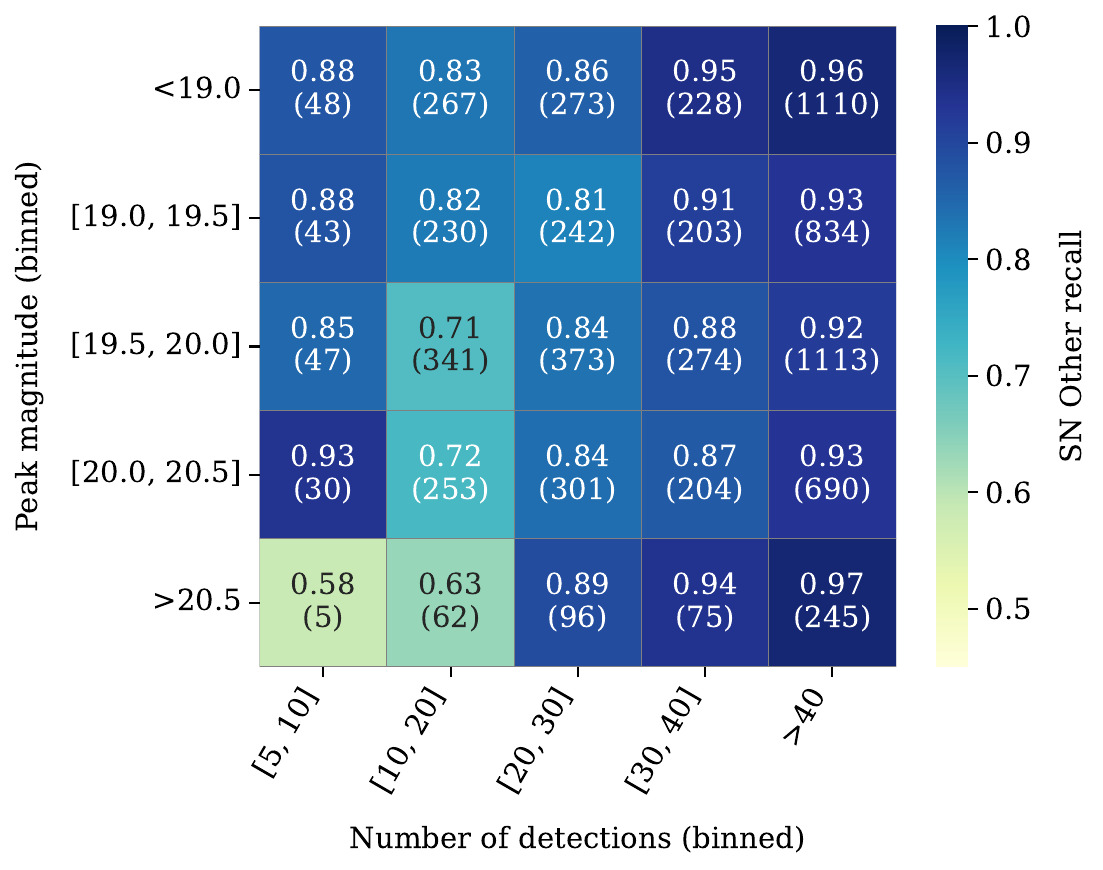}
        \caption{SN~Other recall}
    \end{subfigure}

    \caption{Heatmaps showing the recall of SN~Ia (a) and SN~Other (b) as a function of peak apparent magnitude and number of detections, for the \texttt{RandomZTF+Noise} configuration. Each bin shows the mean recall across five random seeds, with the mean number of objects in the bin indicated in parentheses. We include the mean number of objects per bin to provide a sense of the statistical reliability in each region of parameter space. Note that some bins contain a small number of objects, so their reported recall values should be interpreted with caution. SN~Ia recall remains high (>98\%) for objects brighter than 20 mag and with more than 10 detections, while SN~Other recall improves strongly with increasing number of detections and shows weaker dependence on peak apparent magnitude.}
    \label{figure:mag_vs_det}
\end{figure*}

\subsubsection{Dependence on peak apparent magnitude and number of detections}
\label{subsubsection:results_mag_det_dist}
To better understand the limitations of our classifier and the regimes in which it performs most reliably, we investigated how recall varies as a function of two observational parameters: peak apparent magnitude and number of detections. These parameters characterise the quality and completeness of the light curve, with fainter sources and sparsely sampled light curves generally posing greater challenges for classification.

Figure~\ref{figure:mag_vs_det} shows the heatmaps of SN~Ia and SN~Other recall as functions of peak apparent magnitude and number of detections (in any band), for the \texttt{RandomZTF+Noise} configuration. Each bin displays the mean recall, with the mean number of objects in brackets below. For SNe~Ia, the classifier performs consistently across a wide range of magnitudes and detection counts. SN~Ia recall exceeds 98\% for events with a peak apparent magnitude up to 20 and more than 10 detections, and remains above 96\% up to magnitude 20.5.

In contrast, the SN~Other recall shows a stronger dependence on the number of detections and a weaker dependence on peak apparent magnitude. This indicates that contamination in the SN~Ia sample arises primarily from sparsely sampled light curves rather than from faintness alone. The SN~Other sample is also likely biased toward intrinsically bright sources (e.g., SLSNe) at the faint end of the magnitude bins. This would explain why the recall remains high for faint objects with many detections, as SLSNe are more easily distinguished from SN~Ia than other classes such as SN~Ib/c.

\subsection{Validation with the ZTF SN~Ia DR2 dataset}
\label{subsection:results_dr2}

The ZTF SN~Ia DR2 \citep{Rigault2025} provides a publicly available sample of 3628 spectroscopically classified SNe~Ia discovered by ZTF between March 2018 and December 2020. While much of this sample overlaps with the BTS sample used in training, it also includes additional objects not present in BTS, offering an independent test set for our classifiers.

We applied the same quality cuts as for our test samples (see Sect.~\ref{subsection:create_train}), along with a peak apparent magnitude cut at 18.5 to exclude bright objects that could have appeared in the BTS sample. We also excluded objects that were subtyped as peculiar. Of the 320 remaining SNe~Ia, 96\% were correctly identified by the \texttt{BrightZTF+Noise} classifier, and 97\% were correctly identified by the \texttt{RandomZTF+Noise} classifier. These results are broadly consistent with the SN~Ia recall values reported in Sect.~\ref{subsection:results_best}.

\begin{table}[h!]
\caption{\label{table:dr2_fails} ZTF SN~Ia DR2 sample: photometric classification results. Note that we include misclassifications across all six classifiers, which explains why our correctly classified fraction is lower than the average of 97\%.}
\centering
\begin{tabular}{c p{3.8cm} c}
\hline\hline
Classification & Reason & Fraction (\#)\\ 
\hline
Correct & - & 94.3\% (302) \\
\hline
\multirow{3}{*}{Incorrect} 
 & Bad light curve coverage or bad/failed SALT2 fit & 2.8\% (9) \\
 & Missing photometry due to baseline corrections & 2.2\% (7) \\
 & None of the above & 0.6\% (2) \\ 
\hline
\end{tabular}
\tablefoot{Light curves and spectra for the ZTF SN~Ia DR2 can be found at \href{https://ztfcosmo.in2p3.fr/targetlist}{\texttt{ztfcosmo.in2p3.fr/targetlist}}.}
\end{table}

On average, each classifier failed to select about 10 objects as SNe~Ia. Across the six classifiers (the \texttt{BrightZTF+Noise} model and five random seeds of the \texttt{RandomZTF+Noise} model), 18 unique objects were excluded from the photometrically identified SN~Ia sample. We inspected these objects, and the likely reasons for their misclassification are summarised in Table~\ref{table:dr2_fails}. The majority (2.8\%) were already flagged in the DR2 analysis as having poor light curve coverage (e.g., limited early or late-time data) or poor SALT2 fits (typically due to low fit probabilities derived from the $\chi^2$ distribution). Additionally, seven objects (2.2\%) were missing a significant number of detections in our forced photometry, particularly in the $i$- or $g$-bands, compared to DR2. This is likely due to differences in baseline correction, which can cause detections to be flagged as unreliable and removed. Only two objects (0.6\%) showed no clear reason for misclassification.

\subsection{Discussion of contamination from peculiar subtypes}
\label{subsection:results_contam_pec}
We excluded both peculiar and uncommon subtypes from model training and testing due to insufficient sample sizes to effectively model their behaviour, and to avoid biasing the main classes. However, following the approach of \citet{deSoto2024}, we tested the classifier on these excluded types to examine how they would be classified in practice. The data were processed using the same quality cuts as our test samples (Sect.~\ref{subsection:create_train}). We applied both the \texttt{BrightZTF+Noise} binary and multi-class classifiers to the peculiar test set without augmentation. Table~\ref{table:pec_fails} shows the fraction of each subtype classified as SN~Ia and the fraction misclassified. The total number of objects in each class is shown in parentheses in the left-hand column. Table~\ref{table:pec_multi} expands on this by presenting the predicted class fractions for each peculiar subtype from the multi-class classifier.

\begin{table}[h]
\caption{\label{table:pec_fails} Peculiar objects: SN~Ia classification and misclassification rate.}
\centering
\setlength{\tabcolsep}{4pt}
\begin{tabular}{l|llll}
\hline\hline
True type (\#) & \multicolumn{2}{c}{SN~Ia classification (\%)} & \multicolumn{2}{c}{Misclassified (\%)} \\
 & Binary & Multi & Binary & Multi \\
\hline
Ia-pec (202)  & 48.0 & 39.1 & 52.0 & 60.9 \\
II-pec (134)  & 8.2 & 4.5 & 8.2 & 67.2 \\
Ib/c-pec (34)  & 61.8 & 52.9 & 61.8 & 88.2 \\
TDE (33)  & 6.1 & 6.1 & 6.1 & - \\

\hline
\end{tabular}
\tablefoot{The fractions shown are relative to the true class totals. For the binary classifier, a misclassification occurs when a non-Ia-peculiar object is classified as an SN~Ia. For the multi-class classifier, a misclassification rate is not possible for TDEs since we do not include them in the model training.}
\end{table}

\begin{table}[h]
\caption{\label{table:pec_multi} Peculiar objects: multi-class classifier results.}
\centering
\setlength{\tabcolsep}{4pt}
\begin{tabular}{l|rrrrr}
\hline\hline
True type (\#) & \multicolumn{5}{c}{Predicted class (\%)} \\
 & SN Ia & SN~II & SN~IIn & SN~Ibc & SLSN  \\
\hline
Ia-pec (202) & 39.1 & 4.5 & 6.9 & \textbf{47.0} & 2.4  \\
II-pec (134) & 4.5 & 32.8 & 0.0 & \textbf{62.7} & 0.0 \\
Ib/c-pec (34) & \textbf{52.9} & 26.5 & 8.8 & 11.8 & 0.0  \\
TDE (33)    & 6.1 & 9.1 & 36.3 & 3.0 & \textbf{45.4}\\
\hline
\end{tabular}
\tablefoot{The fractions shown are relative to the true class totals.}
\end{table}

For SN~Ia cosmology, we aim to exclude peculiar SN types from our cosmological sample, which is why our training and test sets only included SNe~Ia, SNe~Ia-norm, and SNe~Ia-91T. However, it is evident that some peculiar subtypes are still classified as SN~Ia by our model. Examining the SN~Ia-pec classifications, we find that most SN~Iax and all SN~Ia-CSM are correctly excluded from the SN~Ia class -- likely because their light curves deviate more significantly from the SALT2 model, with Ia-CSM objects often showing bumps or plateaus from circumstellar interaction. In contrast, large fractions of Ia-91bg, Ia-03fg, Ia-00cx, and Ia-pec are still classified as SN~Ia. To mitigate contamination from these subtypes in a cosmological sample, we could either train a model including these types (potentially using simulations to increase the training sample) or apply additional selection cuts to remove them.

There is significant SN~Ia contamination in the Ib/c-pec class, primarily due to the Ibn subclass. SN~Ibn light curves can exhibit rapid rises and declines that resemble those of some SNe~Ia, making them difficult to distinguish photometrically without spectroscopic data. As with the Ia-pec subtypes, this contamination can be mitigated by applying additional selection cuts to the cosmological sample. SNe~Ibn are also sometimes misclassified as SNe~II because their early blue colours, caused by helium-rich circumstellar interaction, differ significantly from the redder colours of typical SNe~Ib/c and resemble the early phases of SNe~II.

The subtype II-pec, which is mostly comprised of SNe~IIb, is mostly classified as SN~Ib/c. This is expected, since SNe~IIb share similar photometric properties with SNe~Ib/c due to their partially stripped hydrogen envelopes, resulting in similar light curve features such as relatively fast declines compared to SNe~II. Previous studies have shown that the light curve properties of SNe~IIb are observationally distinct from those of SNe~II \citep{Pessi2019}. Based on this, we propose including SNe~IIb in future classifiers either as part of the SN~Ib/c class or as a separate class.

In addition, TDEs are mostly classified as SLSNe or SNe~IIn, likely due to their high luminosity and long-duration light curves. In future work, we aim to include TDEs as a separate class.

\subsection{Application in supernova cosmology}
\label{subsection:results_cosmo}

We have demonstrated that our classifier achieves high accuracy in identifying SNe~Ia, with an efficiency exceeding 98\% for objects brighter than a peak apparent magnitude of 20.5 and having more than 20 detections, when evaluated on real observational data. However, before applying this classifier to the full ZTF sample of unclassified transients, it is essential to simulate the survey realistically. This simulation will allow us to estimate the expected number of SNe~Ia in our sample, quantify contamination from other SN types, estimate selection biases, and assess classifier performance under different survey conditions. This is similar to the approach of the DES collaboration \cite[e.g.,][]{Vincenzi2021,Vincenzi2023}.

When applied to the full ZTF transient sample, our classifier will enable the construction of a large, photometrically classified SN~Ia sample. This dataset will be valuable not only for cosmological analyses, but also for improving the standardisation of SNe~Ia through environmental studies. The Bright Transient Survey in ZTF has recorded approximately 8,000 spectroscopically confirmed SNe~Ia up to a magnitude of 19 by the end of 2024. However, as was shown in the ZTF SN~Ia DR2 \citep{Rigault2025}, many additional SNe~Ia are present in the ZTF sample that were not included in BTS, estimated to number around 10,000. Assuming the number of SNe~Ia roughly doubles with every 0.5 mag of additional depth, we could expect up to 40,000 photometrically classified SNe~Ia up to a magnitude of 21. A key limiting factor remains the availability of redshift information, which is typically obtained from spectroscopic host galaxy surveys. Efforts such as the MOST Hosts Survey \citep{Soumagnac2024} aim to address this by building a comprehensive spectroscopic redshift catalogue for the host galaxies of known transients.

In this work, we prioritised maximising SN~Ia recall to select the best classifier, motivated by previous studies showing that the BBC approach to cosmology with photometric classification can correct for non-Ia contamination \citep[e.g.,][]{Vincenzi2024}. However, \citet{Malz2025} demonstrated that cosmological constraints are sensitive not only to the contamination rate but also to the class of the contaminating population. Traditional classification metrics like recall and precision do not consider the latter. Instead, cosmology-based metrics, such as assessing the cosmological parameter constraints, are more appropriate. We plan to explore these metrics in future studies.

An alternative strategy is proposed by \citet{Boone2021}, in which the same \texttt{ParSNIP} latent-space model is used to jointly classify transients and estimate their distances. Rather than selecting a pure SN~Ia sample, this approach derives distance estimates for all transient types compatible with a given light curve, weighted by their likelihood. This is done by marginalising over the latent-space posterior from the \texttt{ParSNIP} model (i.e., the class probability distribution). The goal is to reduce biases introduced when non-Ia contaminants are assigned distances using models trained on SNe~Ia. Implementing this requires fitting a distance modulus zeropoint model across the full population, including classes with broad luminosity distributions. While promising, this method presents challenges, particularly in accurately modelling the zeropoint and quantifying the light curve modelling uncertainties at the level required for precision cosmology. 

\subsection{The NoiZTF survey: application in real-time classification}
\label{subsection:results_livetest}

Although our primary motivation for developing this classifier was SN~Ia cosmology, it is also effective for real-time transient classification and prioritising candidates for spectroscopic follow-up in ongoing surveys. The relatively nearby ZTF sample provides a unique opportunity to validate our photometric classifications, and we leveraged this by conducting a live follow-up program -- denoted the NoiZTF survey, as in "noisy ZTF" -- as part of the ePESSTO+ collaboration \citep{Smartt2015}.

The ePESSTO+ survey operates on the 3.5-metre NTT at La Silla, using the EFOSC2 instrument to spectroscopically classify transients. We submitted SN-like targets for spectroscopic classification across multiple observing nights between July and November 2024. For this study, we used an earlier version of our classifier that did not include SN~IIn as a separate class; therefore, the performance may not be directly comparable to the results presented in Sect.~\ref{subsection:results_best}. 

Candidates were selected using the \texttt{AMPEL} alert broker \citep{Nordin2019}. We queried the \texttt{AMPEL} API for ZTF alerts that passed basic SN-like criteria from the previous three days. The filtering criteria included a point spread function (PSF) apparent magnitude of greater than $18$; greater than six previous detections but fewer than 50; a real bogus (RB) value of greater than 0.2; and a positive flux value. Probable stars were rejected based on matches to the Pan-STARRS PS1 and Gaia DR2 catalogues. To get estimates for the redshift of the transient, we crossmatched to various galaxy catalogues that were available within \texttt{AMPEL}\footnote{\href{https://ampel.zeuthen.desy.de/api/catalogmatch/docs}{\texttt{ampel.zeuthen.desy.de/api/catalogmatch/docs}}}. The redshift catalogues that we queried are as follows: Sloan Digital Sky Survey (SDSS) DR10 \citep{Brescia2015}; NASA/IPAC Extragalactic Database (NED\footnote{\href{https://ned.ipac.caltech.edu}{\texttt{ned.ipac.caltech.edu}}}), accessed through the \texttt{catsHTM} tool \citep{Soumagnac2018}; Galaxy List for the Advanced Detector Era (GLADE) v2.3 \citep{Dalya2018}; WISExSCOS Photometric Redshift Catalogue \citep[WISExSCOSPZ;][]{Bilicki2016}; 2MASS Photometric Redshift catalogue \citep[2MPZ;][]{Bilicki2014}; Legacy Survey (LS) DR8 \citep{Duncan2022}; Pan-STARRS1 Source Types and Redshifts with Machine Learning \citep[PS1-STRM;][]{Beck2021}.

To exclude old transients or highly variable extragalactic sources (e.g., AGN), we required fewer than 50 detections and a light curve duration of less than 90 days. The multi-class classifier with a probability threshold of 0.7 was applied to the remaining candidate transients. For each candidate, we computed its visibility from La Silla, defined as the number of hours it was observable at an airmass below 2.5, requiring a minimum visibility of one hour. We then assigned priorities to the candidates based on a points-based system, with points awarded for meeting the following criteria:
\begin{enumerate}
    \item The predicted class was rarer (i.e., not SN~Ia),
    \item The object had already peaked (noting this was occasionally inaccurate for sources with re-brightenings or bumps),
    \item The most recent detection was fainter than 19 mag,
    \item The peak apparent magnitude was fainter than 19 mag,
    \item The catalogue-matched redshift was $z < 0.4$ (due to less reliable photometric redshifts at higher $z$).
\end{enumerate}
Candidates were then ranked by total priority score, redshift catalogue reliability, and visibility. While we aimed to select the highest-ranked targets for classification, observational constraints such as scheduling or poor weather at La Silla often required selecting lower-priority (generally brighter) objects.

Table~\ref{table:appendix_noiztfsurvey} in the Appendix lists the 40 sources that received spectroscopic classifications, along with their predicted class, true class, and the number of detections at the time of photometric classification. We also include objects that were submitted to the ePESSTO+ marshall to be observed, but were classified by other groups first and reported on the Transient Name Server (TNS)\footnote{\href{https://www.wis-tns.org/}{\texttt{wis-tns.org}}}. Every candidate submitted was classified as an SN or TDE, except for one case in which the spectrum was too noisy to allow classification. This demonstrates that combining alert brokers with photometric classifiers enables effective prioritisation of real transients for spectroscopic follow-up.

Although the NoiZTF survey was relatively limited in scope (e.g., we were not able to test any SN~Ib/c predictions), it produced several informative results. Of the 19 true SNe~Ia, 17 were correctly classified, giving an SN~Ia recall of 89\%. The two misclassified SNe~Ia were affected by inaccurate high photometric redshifts ($z > 0.4$) from catalogue crossmatches, which likely led to incorrect SLSN classifications. Four non-Ia objects (SNe~II or SNe~Ib/c) were falsely classified as SNe~Ia, resulting in an SN~Ia precision of 81\%. The misclassifications were primarily caused by either sparse light curves -- with fewer than 10 detections -- or, in the case of SN~2024seh, an unusually high luminosity ($M \sim -19.7$) that is atypical for SNe~II.

For the spectroscopically classified SNe~II, 7 out of 11 were correctly identified, resulting in an SN~II recall of 64\%. Misclassifications were again likely caused by inaccurate high photometric redshifts, low detection counts, or intrinsic brightness anomalies (e.g., SN~2024seh). Of the 8 objects predicted by the classifier to be SNe~II, 7 were indeed SNe~II, giving an SN~II precision of 88\%.

The classifier performed particularly well for rare classes such as SLSNe. All four spectroscopically confirmed SLSNe (including one SLSN-IIn) were correctly identified by the classifier (100\% SLSN recall). However, precision was affected by the same redshift errors -- several lower-luminosity objects were incorrectly classified as SLSNe due to inaccurate redshift estimates. One TDE (TDE~2024mvz) was misclassified as an SLSN, which is an expected result given that TDEs were not represented among the classifier’s training classes and TDEs are also known to have high luminosities. Three additional SLSN candidates (SN~2024qef, SN~2024uhx, and AT~2024syj) lacked conclusive spectroscopic classifications due to noisy or featureless spectra. SN~2024qef was initially classified as `SN' due to its blue, featureless spectrum. However, later-phase spectra obtained through a follow-up program with ePESSTO+ were consistent with an SN Ib-pec or Ic-BL classification at $z \sim 0.28$. This redshift implies a peak absolute magnitude of $M < -21$, consistent with an SLSN-I interpretation. An example spectrum and the corresponding spectral template fits are shown in Fig.~\ref{figure:24qef}. AT~2024syj exhibited a light curve lasting over 100 days, consistent with SLSN behaviour. Therefore, a total of 7 out of the 11 objects predicted as SLSNe were either confirmed SLSNe, a TDE, or likely SLSNe, resulting in an effective precision of 64\%.

Across the entire sample -- excluding objects with inconclusive classifications or those not recognised by the classifier (e.g., TDEs) -- we correctly predicted the class for 78\% of the sources. The median number of detections at the time of photometric classification was nine, highlighting the robustness of our method even with limited data. With the latest version of our classifier, this approach provides reliable and efficient target prioritisation for spectroscopic follow-up. This program was a valuable prototype for the Vera C. Rubin Observatory’s LSST, where photometric classification will be essential due to the survey’s depth and data volume. Using real data, we have demonstrated that photometric classification enables efficient selection of promising transient candidates for spectroscopic confirmation, which is vital for the future of time-domain astronomy in LSST.

LSST will provide multi-band photometry (in $ugrizy$), which can significantly aid photometric classification by providing more colour information. This is particularly helpful for distinguishing between SN subtypes with similar light curve shapes but different colours, such as SNe~Ia and SNe~Ib/c. However, LSST’s lower cadence compared to ZTF may limit light curve sampling, posing challenges for accurate classification. In future work, our method will be adapted to LSST photometry, and performance may be further improved by combining our classifier with contextual models that incorporate host galaxy information or are specialised in identifying rarer classes, such as TDEs. A hybrid framework that combines multiple classifiers will likely be essential to enable reliable large-scale classification in the LSST era.


\section{Conclusions}
\label{section:conc}
In this work, we presented a feature-based photometric classifier for SNe detected by ZTF, with the primary goal of constructing a photometric SN~Ia sample for cosmological analyses. Our approach utilises the autoencoder architecture from the \texttt{ParSNIP} model \citep{Boone2021} to capture the intrinsic diversity of SN light curves. We trained the model on the ZTF SN sample, incorporating a realistic noise augmentation procedure that simulates the flux uncertainties of fainter sources. This enables the model to generalise to noisier, higher-redshift populations. Light curve features were used to train a gradient-boosted decision tree classifier, implemented in both binary (SN~Ia vs. non-Ia) and multi-class configurations. We validated our classifier on independent, fainter ZTF data with and without noise augmentation. To evaluate real-time performance, we also applied our classifier to live ZTF alerts and conducted a spectroscopic classification survey within the ePESSTO+ collaboration.

The conclusions of our analysis are as follows:
\begin{enumerate}
    \item Noise augmentation significantly improves classification performance, particularly for fainter sources. We showed that SN~Ia recall increases from 90\% to 95\% on real data, and from 86\% to 96\% on fainter, noise-augmented data.
    \item Our binary classifier achieves an SN~Ia recall of $(98.1 \pm 0.4)\%$, averaged across five train-test splits. The classifier performs consistently across a wide range of magnitudes and detection counts: SN~Ia recall exceeds 98\% for events with a peak apparent magnitude up to 20 and more than 10 detections, and remains above 96\% up to magnitude 20.5. 
    \item Overall, 95\% of sources were correctly classified in both binary and multi-class modes.
    \item  In our live classification survey, we correctly identified the class for 78\% of the targets, including rare events such as SLSNe, despite a median of only nine detections per object. We measured an SN~Ia recall of 89\% and a precision of 81\%. All four spectroscopically confirmed SLSNe were correctly identified by the classifier (100\% SLSN recall). Of the 11 objects predicted as SLSNe, 7 were either confirmed SLSNe, a TDE, or likely SLSNe, resulting in an effective precision of 64\%. This is notable given that the classification relied on photometric redshift catalogues to estimate the transient redshift.
\end{enumerate}

Our classifier performs efficiently on real ZTF data, including faint and noisy light curves. When applied to the full ZTF transient sample, our classifier will enable the construction of a large photometrically classified SN~Ia sample for cosmology. Our next steps include simulating the ZTF survey realistically. This simulation will allow us to estimate the expected number of SNe~Ia in our photometric sample, quantify contamination from other SN types, estimate selection biases, and assess classifier performance under survey conditions. Additionally, our methodology offers an effective tool for real-time target prioritisation for spectroscopic follow-up, which will be essential for future large-scale surveys such as LSST.

\section{Data availability}
\label{section:datavail}
ZTF light curve data is publicly available at \href{https://www.ztf.caltech.edu/ztf-public-releases.html}{\texttt{ztf.caltech.edu/ztf-public-releases.html}}. A repository that contains the code used to generate the training and test samples, along with the model files for feature generation from any light curve, and the classifier files for running your own classifications can be found here: \href{https://github.com/aotownsend/noiztf/tree/v1.0.0}{\texttt{github.com/aotownsend/noiztf/tree/v1.0.0}}.

\begin{acknowledgements}
Based on observations obtained with the Samuel Oschin Telescope 48-inch and the 60-inch Telescope at the Palomar Observatory as part of the Zwicky Transient Facility project. ZTF is supported by the National Science Foundation under Grants No. AST-1440341, AST-2034437, and currently Award \#2407588. ZTF receives additional funding from the ZTF partnership. Current members include Caltech, USA; Caltech/IPAC, USA; University of Maryland, USA; University of California, Berkeley, USA; University of Wisconsin at Milwaukee, USA; Cornell University, USA; Drexel University, USA; University of North Carolina at Chapel Hill, USA; Institute of Science and Technology, Austria; National Central University, Taiwan, and OKC, University of Stockholm, Sweden. Operations are conducted by Caltech's Optical Observatory (COO), Caltech/IPAC, and the University of Washington at Seattle, USA. The ZTF forced-photometry service was funded under the Heising-Simons Foundation grant \#12540303 (PI: Graham). The Gordon and Betty Moore Foundation, through both the Data-Driven Investigator Program and a dedicated grant, provided critical funding for SkyPortal.\\

Based on observations collected at the  European Organisation for Astronomical Research in the Southern Hemisphere, Chile, as part of ePESSTO+ (the advanced Public ESO Spectroscopic Survey for Transient Objects Survey – PI: Inserra), under ESO program ID 112.25JQ. \\

K.M., T.E.M.B. and U.B. acknowledge funding from Horizon Europe ERC grant no. 101125877. T.-W.C. acknowledges the financial support from the Yushan Fellow Program by the Ministry of Education, Taiwan (MOE-111-YSFMS-0008-001-P1) and the National Science and Technology Council, Taiwan (NSTC grant 114-2112-M-008-021-MY3). Y.-L.K. was supported by the Lee Wonchul Fellowship, funded through the BK21 Fostering Outstanding Universities for Research (FOUR) Program (grant No. 4120200513819) and the National Research Foundation of Korea to the Center for Galaxy Evolution Research (RS-2022-NR070872, RS-2022-NR070525). S.Y. acknowledges the funding from the National Natural Science Foundation of China under grant No. 12303046, and the Startup Research Fund of Henan Academy of Sciences No.242041217. C.P.G. acknowledges financial support from the Secretary of Universities and Research (Government of Catalonia) and by the Horizon 2020 Research and Innovation Programme of the European Union under the Marie Sk\l{}odowska-Curie and the Beatriu de Pin\'os 2021 BP 00168 programme, from the Spanish Ministerio de Ciencia e Innovaci\'on (MCIN) and the Agencia Estatal de Investigaci\'on (AEI) 10.13039/501100011033 under the PID2023-151307NB-I00 SNNEXT project, from Centro Superior de Investigaciones Cient\'ificas (CSIC) under the PIE project 20215AT016 and the program Unidad de Excelencia Mar\'ia de Maeztu CEX2020-001058-M, and from the Departament de Recerca i Universitats de la Generalitat de Catalunya through the 2021-SGR-01270 grant. L.G. acknowledges financial support from AGAUR, CSIC, MCIN and AEI 10.13039/501100011033 under projects PID2023-151307NB-I00, PIE 20215AT016, CEX2020-001058-M, ILINK23001, COOPB2304, and 2021-SGR-01270.

\end{acknowledgements}

\bibliographystyle{aa}
\bibliography{phot.bib}

\begin{appendix} 

\onecolumn
\section{K-correction spectral templates}

Section~\ref{subsection:kcorr} details the method used to compute K-corrections for our augmented samples. Table~\ref{table:appendix_kcorr} lists the \texttt{sncosmo} template models employed in this process.

\begin{table}[h]
\caption{\label{table:appendix_kcorr} \texttt{sncosmo} templates for K-corrections}
\centering
\begin{tabular}{ll}
\hline\hline
SN type & \texttt{sncosmo} template\\ 
\hline
SN Ia and subtypes & salt2 \\
SN Ib, Ib-pec and Ibn & v19-iptf13bvn-corr\\
SN Ib/c & v19-iptf13bvn-corr\\
SN Ic and Icn & v19-2013ge-corr\\
SN Ic-BL & v19-2012ap-corr\\
SN II and II-pec & v19-2016bkv-corr\\
SN IIb and IIb-pec & v19-2016gkg-corr \\
SN IIL & s11-2004hx\\
SN IIP & snana-2007iz\\
SN IIn & v19-2011ht-corr\\
\hline
\end{tabular}
\tablefoot{The relevant references for each spectral template are found at \href{https://sncosmo.readthedocs.io/en/stable/source-list.html}{\texttt{sncosmo.readthedocs.io/en/stable/source-list.html}}.}
\end{table}

\section{Example augmented light curves}
\begin{figure*}[h!]
\centering
\includegraphics[width=\hsize]{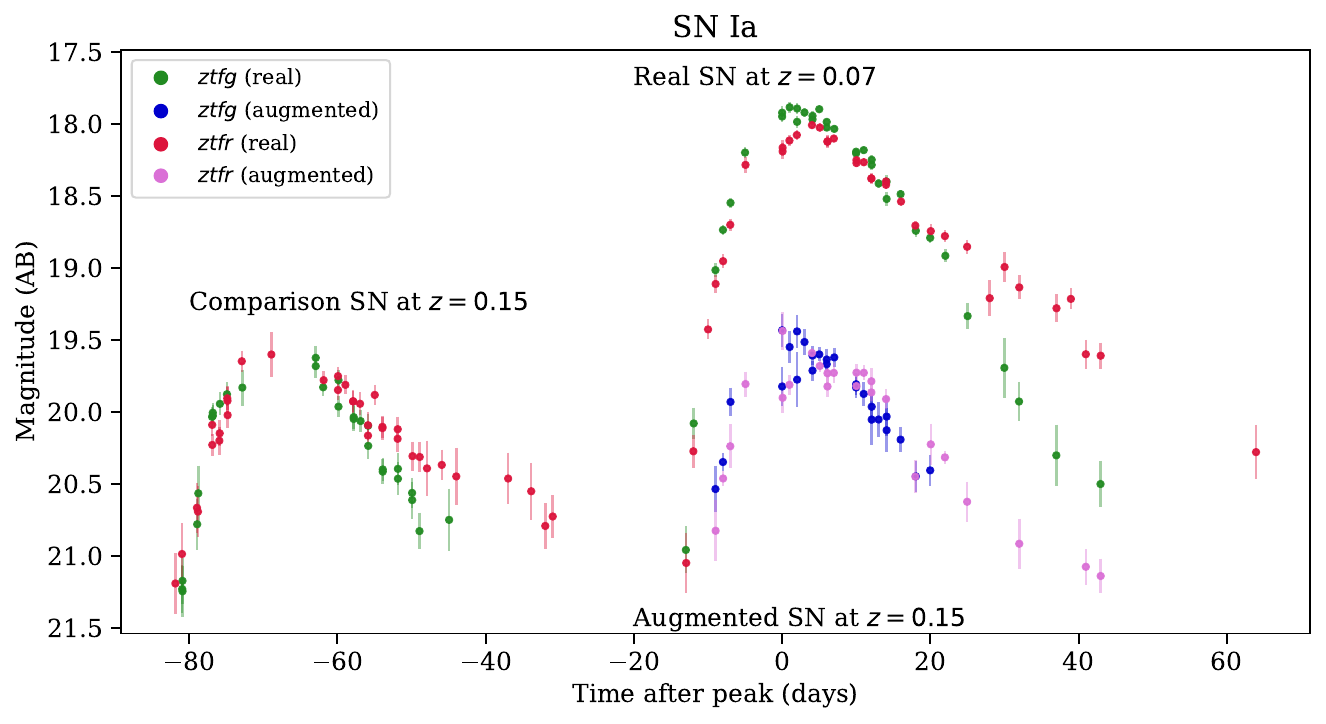}
  \caption{Comparison of a real, bright SN~Ia (SN~2021vts at $z=0.072$), its augmented counterpart at higher redshift ($z=0.147$), and a real SN~Ia at a similar redshift (SN~2021kfy at $z=0.15$). Data points are shown in the $g$-band (green for real, pink for augmented) and $r$-band (red for real, blue for augmented); the $i$-band is omitted for clarity. Time is shown relative to the peak of SN~2021vts, with an offset applied to SN~2021kfy to aid comparison.
      }
 \label{figure:example_Ia}
 \end{figure*}

\begin{figure*}[h!]
\centering
\includegraphics[width=\hsize]{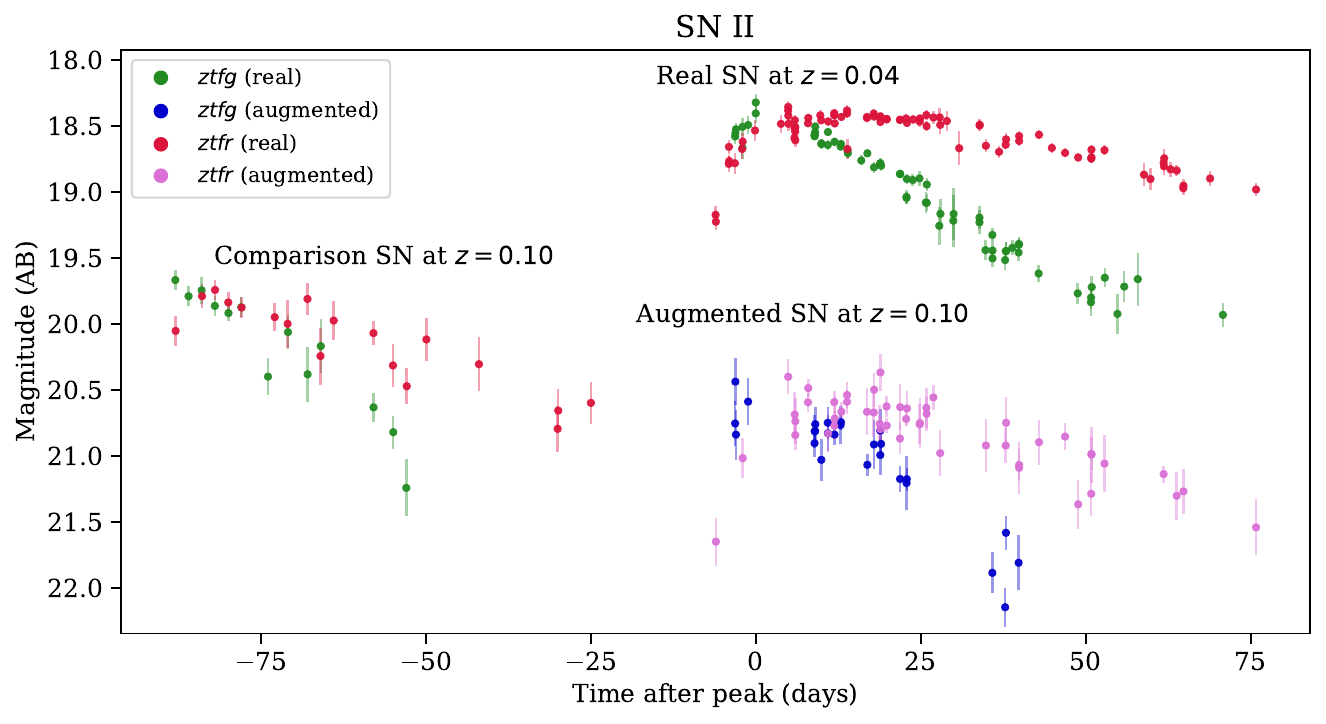}
  \caption{Comparison of a real, bright SN~II (SN~2022jnh at $z=0.038$), its augmented counterpart at higher redshift ($z=0.095$), and a real SN~II at a similar redshift (SN~2020whv at $z=0.097$). Data points are shown in the $g$-band (green for real, pink for augmented) and $r$-band (red for real, blue for augmented); the $i$-band is omitted for clarity. Time is shown relative to the peak of SN~2022jnh, with an offset applied to SN~2020whv to aid comparison. Note that, unlike SNe~Ia, SNe~II have more diverse peak luminosities, so the augmented and comparison light curves peak at different magnitudes.
      }
 \label{figure:example_II}
 \end{figure*}

Section~\ref{section:method} details the procedure used to generate the augmented light curves used in our training and test samples. Figures~\ref{figure:example_Ia} and \ref{figure:example_II} present examples of augmented SN~Ia and SN~II light curves, respectively. For comparison, each panel includes a real light curve at a similar redshift, illustrating that the augmented data exhibit realistic brightness evolution and scatter consistent with real observations. The augmented light curves shown were drawn from the training sample, generated using the augmentation parameters: $\mathrm{\texttt{z\_scale}}=0,\: \mathrm{\texttt{cadence\_scale}}=100,\: \mathrm{\texttt{subsampling\_rate}}=0.9$.

\section{Augmentation parameter optimisation: \texttt{cadence\_scale} and \texttt{subsampling\_rate}}

\begin{figure}[h!]
\centering
\includegraphics[width=0.5\hsize]{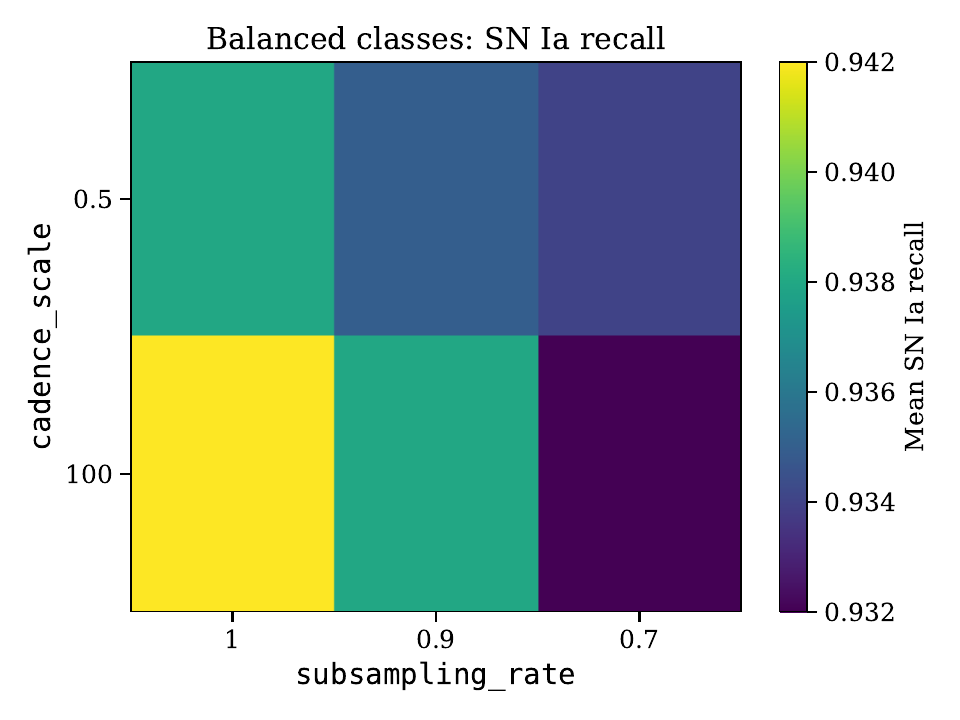}
  \caption{Mean SN~Ia recall, averaged over multiple iterations of the \texttt{FaintZTF} and \texttt{FaintZTF+Noise} test sets (with balanced class scaling), as a function of the augmentation parameters \texttt{cadence\_scale} and \texttt{subsampling\_rate}. The highest recall occurs when no data points are removed, corresponding to \texttt{cadence\_scale} = 100 and \texttt{subsampling\_rate} = 1.
      }
 \label{figure:cadencevssubsamp_Ia}
 \end{figure}

\begin{figure}[h!]
\centering
\includegraphics[width=0.5\hsize]{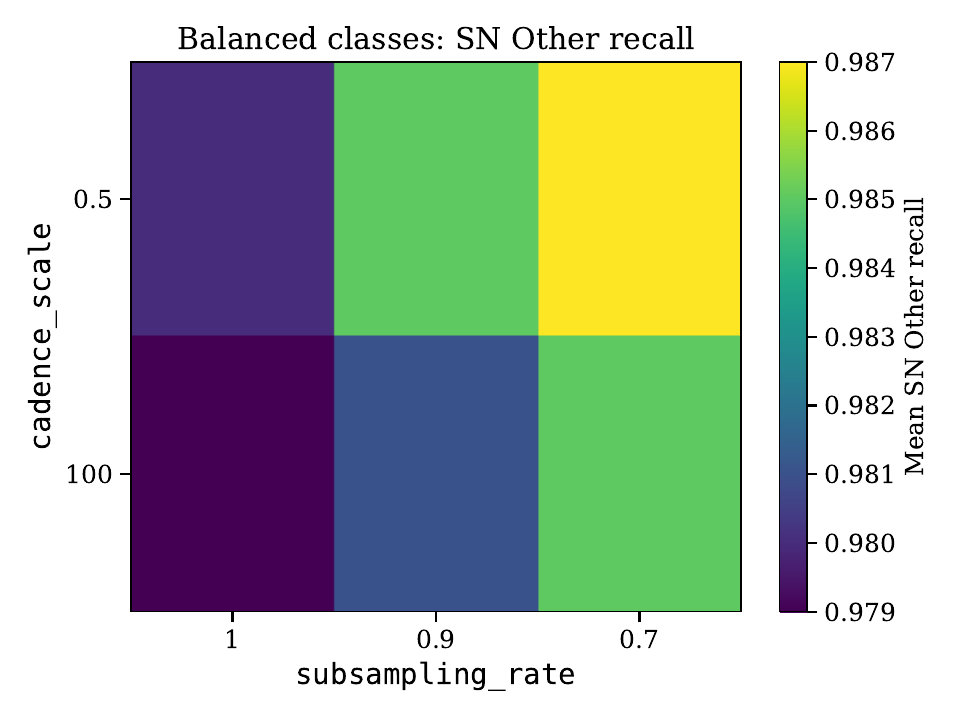}
  \caption{Mean SN~Other recall, averaged over multiple iterations of the \texttt{FaintZTF} and \texttt{FaintZTF+Noise} test sets (with balanced class scaling), as a function of the augmentation parameters \texttt{cadence\_scale} and \texttt{subsampling\_rate}. Performance shows a slight preference for more aggressive data removal, with optimal values around \texttt{cadence\_scale} = 0.5 and \texttt{subsampling\_rate} = 0.7.
      }
 \label{figure:cadencevssubsamp_nonIa}
 \end{figure}

Section~\ref{subsection:results_param_opt} outlines the augmentation parameters we tested to optimise classifier performance. While class scaling had the most significant impact, we also observed a weaker dependence on the \texttt{cadence\_scale} and \texttt{subsampling\_rate} parameters. As shown in Figs.~\ref{figure:cadencevssubsamp_Ia} and \ref{figure:cadencevssubsamp_nonIa}, SN~Ia recall was highest when no data points were removed, whereas SN~Other recall improved with more aggressive subsampling.

\onecolumn
\section{Confusion matrices for the \texttt{BrightZTF} and \texttt{FaintZTF} training and test samples}
\begin{figure*}[htbp]
    \centering
    \begin{subfigure}[b]{0.4\textwidth}
        \centering
        \caption{Train: \texttt{BrightZTF}, Test: \texttt{FaintZTF}}
        \includegraphics[width=\linewidth]{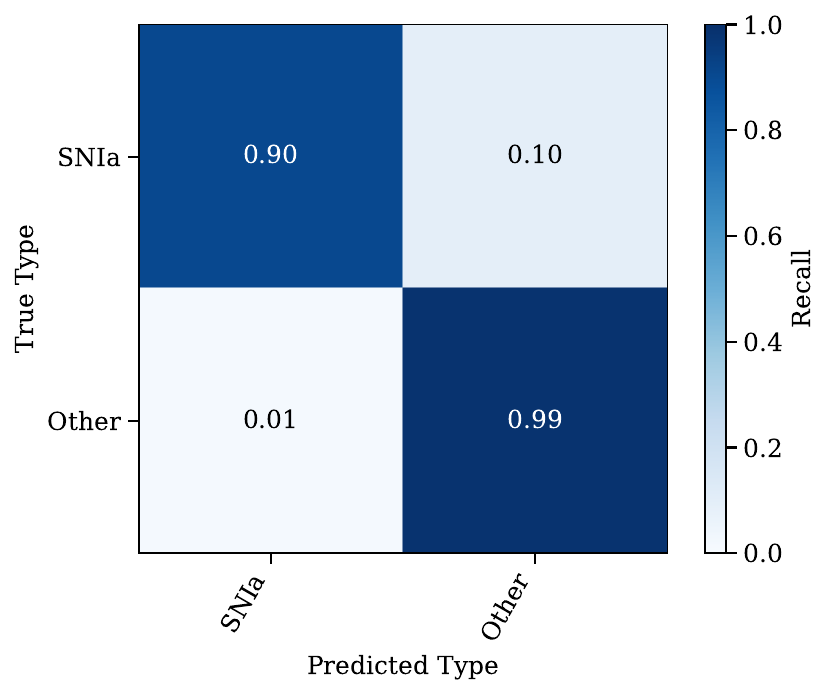}
    \end{subfigure}
    \begin{subfigure}[b]{0.4\textwidth}
        \centering
        \caption{Train: \texttt{BrightZTF}, Test: \texttt{FaintZTF+Noise}}
        \includegraphics[width=\linewidth]{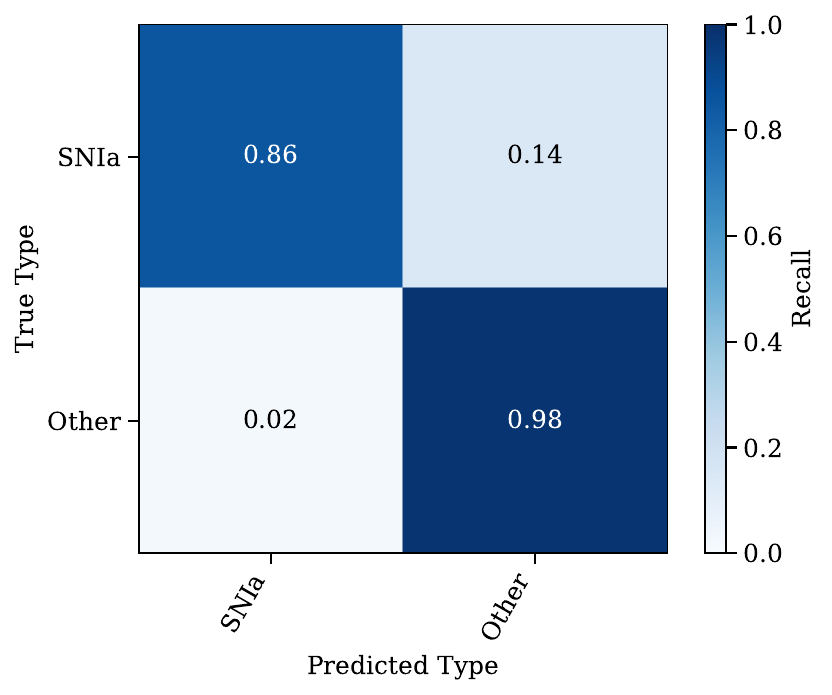}
    \end{subfigure}
    
    \vskip\baselineskip
    
    \begin{subfigure}[b]{0.4\textwidth}
        \centering
        \caption{Train: \texttt{BrightZTF+Noise}, Test: \texttt{FaintZTF}}
        \includegraphics[width=\linewidth]{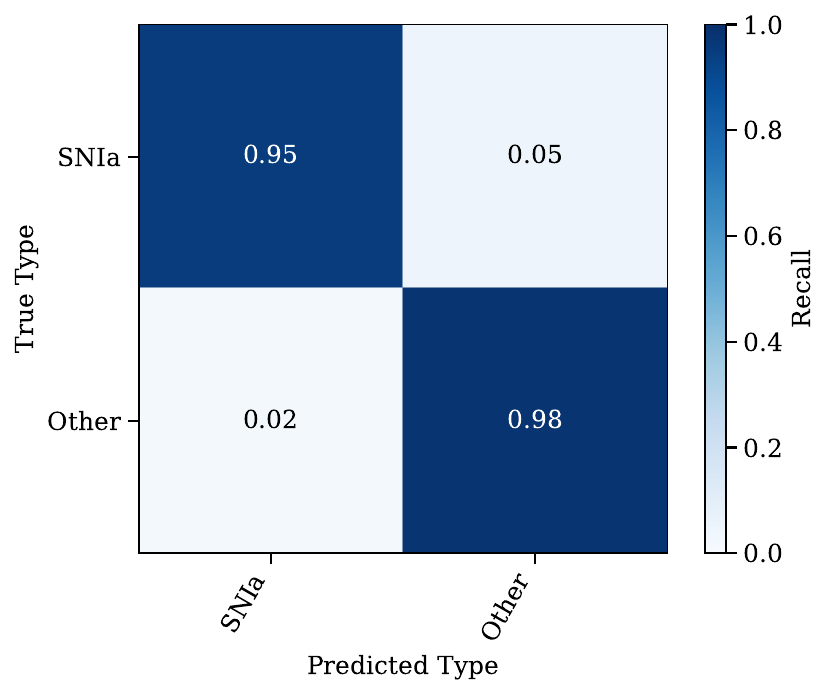}
    \end{subfigure}
    \begin{subfigure}[b]{0.4\textwidth}
        \centering
        \caption{Train: \texttt{BrightZTF+Noise}, Test: \texttt{FaintZTF+Noise}}
        \includegraphics[width=\linewidth]{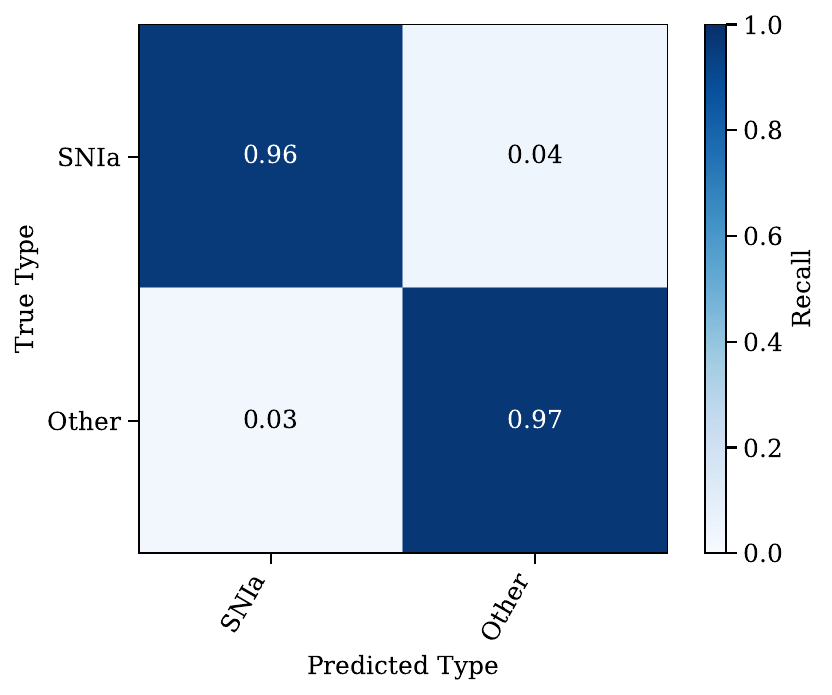}
    \end{subfigure}

    \caption{Confusion matrices for our selected model configuration (bolded in Table~\ref{table:aug_params}), normalised by true class totals (i.e., by row) to display the recall across different augmentation combinations of \texttt{BrightZTF} and \texttt{FaintZTF}. The four plots correspond to: (a) training and testing on unaugmented data, (b) training on unaugmented data and testing on augmented data, (c) training on augmented data and testing on unaugmented data, and (d) training and testing on augmented data. The results demonstrate that noise augmentation in the training set improves classification performance on fainter, noisier test samples.}
    \label{figure:brightfaint_cm}
\end{figure*}

We discuss in Sect.~\ref{subsection:results_best} how SN~Ia recall improves with augmentation in the training sample, resulting in better classification of fainter, noisier light curves. Figure~\ref{figure:brightfaint_cm} shows the confusion matrices, normalised by true class totals (i.e., by row), to display the recall across different augmentation combinations of \texttt{BrightZTF} and \texttt{FaintZTF}.

\FloatBarrier
\newpage

\section{The NoiZTF survey: classification results}
In Sect.~\ref{subsection:results_livetest}, we presented the NoiZTF survey, a spectroscopic follow-up program designed to select interesting candidates and evaluate the photometric classifications produced by our classifier. Table~\ref{table:appendix_noiztfsurvey} summarises the results from this classification survey.

\begin{figure}[h!]
\centering
\includegraphics[width=0.6\hsize]{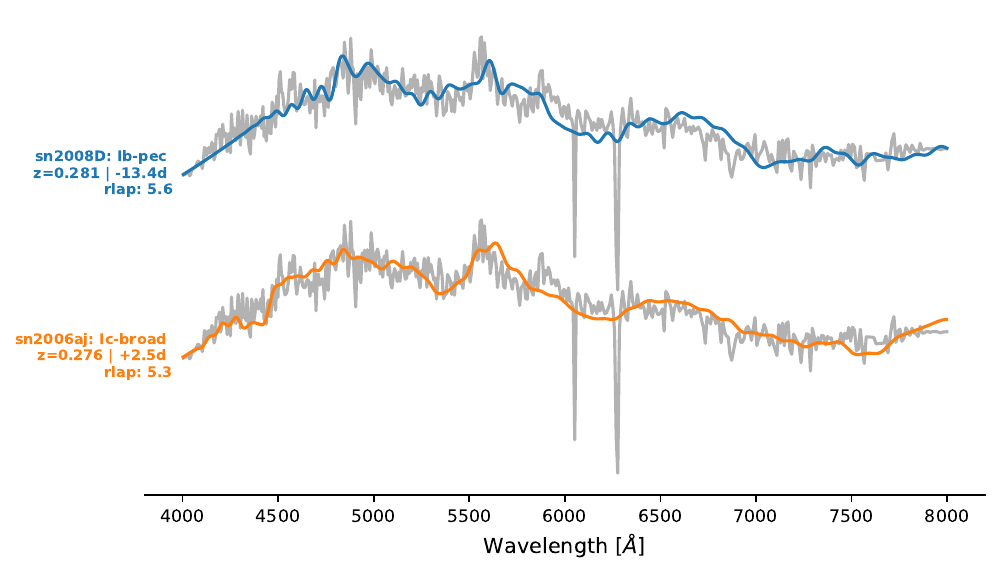}
  \caption{A late-phase spectrum of SN~2024qef, taken on 2 October 2024 by the ePESSTO+ collaboration as part of a follow-up program, approximately 80 days after the first detection. The two best-matching spectral templates correspond to a peculiar Type Ib (blue) and a broad-lined Type Ic (orange), both at a redshift of $z \sim 0.28$.
      }
 \label{figure:24qef}
 \end{figure}

Figure~\ref{figure:24qef} shows a late-phase spectrum of SN~2024qef, taken on 2 October 2024 by the ePESSTO+ collaboration as part of a follow-up program, approximately 80 days after the first detection. The data was reduced\footnote{\href{https://github.com/svalenti/pessto}{\texttt{github.com/svalenti/pessto}}}, and spectral template fitting was performed using the python wrapper \texttt{pysnid}\footnote{\href{https://github.com/MickaelRigault/pysnid}{\texttt{github.com/MickaelRigault/pysnid}}} of the Supernova Identification code \citep[\texttt{SNID};][]{Blondin2007}\footnote{\href{https://people.lam.fr/blondin.stephane/software/snid/}{\texttt{people.lam.fr/blondin.stephane/software/snid/}}}. The two best-matching templates correspond to a peculiar Type Ib and a broad-lined Type Ic, both at a redshift of $z \sim 0.28$. Combining this redshift with the photometric data (peak apparent magnitude $r_{ZTF} \sim 19.3$) results in a peak absolute magnitude of $M < -21$, which is consistent with an SLSN-I interpretation. If this interpretation is correct, it suggests that the photometric classification was more informative than the early-phase spectroscopy, which produced only a blue, featureless spectrum. This highlights the potential of photometric classifiers to provide meaningful classifications in cases where spectroscopy is inconclusive -- particularly during the early phases, when certain transients are intrinsically featureless.

\begin{table}[h]
\caption{\label{table:appendix_noiztfsurvey} NoiZTF classification survey results.}
\centering
\begin{tabular}{llllll}
\hline\hline
    ZTF name &   IAU name & Predicted class & True class &  No. of detections & Reference \\
\hline
ZTF24aaooxcd\textsuperscript{ 1} &  SN~2024jfh &            SN~II &       SN~II &                 33 & \citet{20240712_TNS} \\
ZTF24aarqqug &  SN~2024lhf &            SN~Ia &       SN~Ia &                 27 & \citet{20240630_TNS} \\
ZTF24aardmeg &  SN~2024lcm &            SN~Ia &       SN~Ia &                  8 & \citet{20240712_TNS} \\
ZTF24aapzmaf &  SN~2024kng &            SN~Ia &       SN~Ia &                 13 & \citet{20240712_TNS} \\
ZTF24aaowjlw &  SN~2024jwx &            SLSN &       SLSN &                 33 & \citet{20240705_TNS} \\
ZTF24aarqrkn\textsuperscript{ 2} &  SN~2024kzu &            SLSN &       SN~Ia &                 24 & \citet{20240711_TNS} \\
ZTF24aarkrxi &  SN~2024lgz &            SN~II &       SN~II &                 29 & \citet{20240712_TNS} \\
ZTF24aaumhmm &  SN~2024nzi &            SN~Ia &       SN~Ia &                  7 & \citet{20240712_TNS} \\
ZTF24aauerna &  SN~2024nzf &            SN~Ia &       SN~Ia &                 17 & \citet{20240713_TNS} \\
ZTF24aaumgko &  SN~2024nzg &            SN~Ia &       SN~Ia &                  5 & \citet{20240713_TNS} \\
ZTF24aashinc &  SN~2024ltj &            SN~Ia &       SN~Ia &                 13 & \citet{20240711_TNS} \\
ZTF24aaseunp &  SN~2024lrp &            SN~Ia &       SN~Ia &                  7 & \citet{20240711_TNS} \\
ZTF24aaxhvbu\textsuperscript{ 3} &  SN~2024qef &            SLSN &         SN &                 29 & \citet{20240831_TNS} \\
ZTF24aaysowl &  SN~2024rmj &            SLSN &       SLSN &                  9 & \citet{20240902_TNS}\\
ZTF24abbjtjc &  SN~2024ser &            SN~Ia &       SN~Ia &                 13 & \citet{20240902_TNS} \\
ZTF24aasteui\textsuperscript{ 4} &  TDE~2024mvz &            SLSN &        TDE &                 16 & \citet{20240902_TNS} \\
ZTF24aatsmxc &  SN~2024nbv &            SN~II &       SN~II &                 24 & \citet{20240902b_TNS}  \\
ZTF24abbcrsy\textsuperscript{ 2} &  SN~2024sbx &            SLSN &       SN~II &                  9 & \citet{20240828_TNS} \\
ZTF24abavoif\textsuperscript{ 5} &  SN~2024seh &            SN~Ia &       SN~II &                 15 & \citet{2024TNS_Mehta} \\
ZTF24abcnnlb &  SN~2024snk &            SN~Ia &       SN~Ia &                  6 & \citet{20240827_TNS} \\
ZTF24abbsuhu\textsuperscript{ 6} &  SN~2024sjw &            SN~Ia &      SN~Ib/c &                  7 & \citet{20240902b_TNS} \\
ZTF24abbggnw &  SN~2024nrw &            SLSN &       SLSN &                  7 & \citet{20240831_TNS} \\
ZTF24abayoiy &  SN~2024sjd &            SN~Ia &       SN~Ia &                  5 & \citet{20240902_TNS} \\
ZTF24abeiasc\textsuperscript{ 7} &  SN~2024uhx &            SLSN &         SN &                 13 & \citet{20240909_TNS} \\
ZTF24abamahv\textsuperscript{ 8} &  AT~2024syj &            SLSN &          - &                 12 & \citet{20241003_TNS} \\
ZTF24aaxqoeo &  SN~2024pll &            SN~II &       SN~II &                  7 & \citet{20240907_TNS} \\
ZTF24abojqpw &  SN~2024zgv &            SN~Ia &       SN~Ia &                  8 & \citet{20241106_TNS} \\
ZTF24aboiwjo\textsuperscript{ 2} &  SN~2024zjt &            SLSN &       SN~Ia &                  7 & \citet{20241110_TNS} \\
ZTF24abiyona &  SN~2024uki &            SN~II &       SN~II &                 15 & \citet{20240909_TNS} \\
ZTF24abrddqo & SN~2024aaqc &            SN~Ia &       SN~Ia &                  6 & \citet{20241108_TNS} \\
ZTF24abpdzvm\textsuperscript{ 9,10} &  SN~2024zsw &            SN~II &       SN~II &                 19 & \citet{2024TNS_YSE} \\
ZTF24abpvane &  SN~2024zwm &            SN~Ia &       SN~Ia &                  8 & \citet{20241112_TNS} \\
ZTF24abrkfol\textsuperscript{10} & SN~2024abbi &            SN~Ia &       SN~Ia &                  6 & \citet{2024TNS_SollermanA} \\
ZTF24abtbalq & SN~2024abop &            SN~Ia &       SN~II &                  9 & \citet{20241122_TNS} \\
ZTF24abrfiya & SN~2024aasm &            SN~Ia &       SN~II &                  9 & \citet{20241130_TNS} \\
ZTF24abrfhyl\textsuperscript{ 10} & SN~2024aaqs &            SN~II &       SN~II &                  8 & \citet{2024TNS_Blagorodnova} \\
ZTF24abqbyxd\textsuperscript{ 10} &  SN~2024zzf &            SN~II &      SN~Ib/c &                  9 & \citet{2024TNS_Fremling} \\
ZTF24abpiwwi\textsuperscript{ 9,11} & SN~2024aaks &            SLSN &      SLSN-IIn &                 40  & \citet{2024TNS_Angus}  \\
ZTF24abtcetf\textsuperscript{ 10} & SN~2024abnt &            SN~Ia &       SN~Ia &                  7 & \citet{2024TNS_SollermanB} \\
ZTF24abubygj\textsuperscript{ 10} & SN~2024acon &            SN~Ia &       SN~Ia &                  6 & \citet{2024TNS_Covarrubias} \\
\hline
\end{tabular}
\begin{flushleft}
\small
\textsuperscript{1} SN~2024jfh was initially photometrically classified as an SN~Ia with $p > 0.7$ at 23 detections, which changed to an SN~II with $p=0.65$ at 33 detections. \\
\textsuperscript{2} These objects were likely photometrically classified as SLSNe due to an erroneous high photometric redshift ($z > 0.4$) from a catalogue crossmatch.\\
\textsuperscript{3} SN~2024qef was classified as an SN with a blue featureless spectrum. Follow-up spectra from ePESSTO+ taken at later phases have good matches to Ib-pec or Ic-BL at $z \sim0.28$. This would mean a peak absolute magnitude of $M<-21$ mag, suggesting the true type could be an SLSN-I. \\
\textsuperscript{4} Given that the TDE was not available to the classifier, a prediction of SLSN is the most likely outcome. \\
\textsuperscript{5} SN~2024seh was initially classified as SN by ePESSTO+, and then further classified as SN~II by another group.\\
\textsuperscript{6} SN~2024sjw was classified as an SN~Ib. \\
\textsuperscript{7} SN~2024uhx was classified as an SN with a blue featureless spectrum. \\
\textsuperscript{8} AT~2024syj was observed, but the spectrum was too noisy for a conclusive classification. However, its light curve duration of over 100 days is consistent with an SLSN. \\
\textsuperscript{9} SN~2024zsw was further classified as an SN~IIb by a Young Supernova Experiment group. \\
\textsuperscript{10} Not classified by ePESSTO+, but by a ZTF groups reporting to TNS.\\
\textsuperscript{11} Classified by the SCAT group as an SN~IIn. However, the peak absolute magnitude of this object is $M<-21$ mag, suggesting it is an SLSN-IIn.
\end{flushleft}
\tablefoot{The number of detections refers to those available at the time of photometric classification, not at the time of spectroscopic classification. Classifications and spectra are publicly available at: \href{https://www.wis-tns.org/}{\texttt{wis-tns.org}}.}
\end{table}

\end{appendix}

\end{document}